\documentclass[12pt]{iopart}
\usepackage{epsfig}
\usepackage{iopams}

\newcommand{\vc}[1]{\bi{#1}}
\newcommand{\mx}[1]{\mathbf{#1}}
\renewcommand{\d}{\rmd}
\newcommand{\pp}[2][]{\ensuremath{\frac{\partial #1}{\partial #2}}}
\DeclareMathSymbol{\mg}{\mathrel}{symbols}{"1D}  
\newcommand{\Bnabla}{\bnabla}
\newcommand{\Bder}{\bpartial}
\newcommand{\half}{\frac 12 }
\newcommand{\Id}{\mbox{\small 1}\hspace{-3.5pt}\mbox{1}}
\newcommand{\lra}{\longrightarrow}
\newcommand{\Ra}{\Rightarrow}
\newcommand{\ra}{\rightarrow}
\newcommand{\der}{\partial}
\newcommand{\inv}{^{-1}}
\newcommand{\lh}{\left(}
\newcommand{\rh}{\right)}
\newcommand{\labl}[1]{\label{#1}}
\newcommand{\eqref}[1]{(\ref{#1})}
\newcommand{\Dot}[1]{\dot{#1}}
\renewenvironment{matrix}{%
  \hskip -\arraycolsep\array{cccccccccc}%
}{%
  \endarray \hskip -\arraycolsep
}
\renewenvironment{pmatrix}{\left(\matrix}{\endmatrix\right)}
\newcommand{\equ}[1]{\begin{eqnarray} #1 \end{eqnarray}}
\newcommand{\items}[1]{\begin{itemize} #1 \end{itemize}}
\newcommand{\enums}[1]{\begin{enumerate} #1 \end{enumerate}}
\newcommand{\tabu}[2]{\begin{tabular}{#1} #2 \end{tabular}}
\newcommand{\pmtrx}[1]{\begin{pmatrix} #1 \end{pmatrix}}
\newcommand{\non}{\nonumber}

\newcommand{\Real}{\mathbb{R}}
                               
\newcommand{\Bsfm}{\mbox{{\sffamily\bfseries m}}}   
\newcommand{\ga}{\alpha}
\newcommand{\gb}{\beta}
\renewcommand{\gg}{\gamma}
\newcommand{\gd}{\delta}
\renewcommand{\ge}{\epsilon}

\newcommand{\gf}{\phi}

\newcommand{\gx}{\xi}
\newcommand{\gm}{\mu}
\newcommand{\gn}{\nu}

\newcommand{\gk}{\kappa}
\newcommand{\gl}{\lambda}
\newcommand{\gr}{\rho}
\newcommand{\gth}{\theta}

\newcommand{\gt}{\tau}
\newcommand{\go}{\omega}
\newcommand{\gz}{\zeta}
\newcommand{\gp}{\pi}
\newcommand{\gps}{\psi}
\newcommand{\get}{\eta}

\newcommand{\gG}{\Gamma}
\newcommand{\gD}{\Delta}
\newcommand{\gF}{\Phi}

\newcommand{\gO}{\Omega}
\newcommand{\gP}{\Pi}

\newcommand{\cD}{{\cal D}}
\newcommand{\cH}{{\cal H}}
\newcommand{\cL}{{\cal L}}
\newcommand{\cM}{{\cal M}}
\newcommand{\cO}{{\cal O}}
\newcommand{\ta}{{\tilde a}}
\newcommand{\tb}{{\tilde b}}

\newcommand{\tC}{{\tilde C}}
\newcommand{\tD}{{\tilde D}}

\newcommand{\tM}{{\tilde M}}

\newcommand{\tQ}{{\tilde Q}}

\newcommand{\tS}{{\tilde S}}

\newcommand{\tU}{{\tilde U}}

\newcommand{\bz}{{\bar z}}
\newcommand{\bQ}{{\bar Q}}
\newcommand{\bga}{{\bar\alpha}}
\newcommand{\bgO}{{\bar\Omega}}
\newcommand{\uga}{{\underline\alpha}}
\newcommand{\tge}{{\tilde\epsilon}}

\newcommand{\tgx}{{\tilde\xi}}

\newcommand{\tget}{{\tilde\eta}}

\newcommand{\tgO}{{\tilde\Omega}}
\newcommand{\Bgd}{\bdelta}

\newcommand{\Bgf}{\bphi}

\newcommand{\Bgx}{\bxi}

\newcommand{\Bget}{\bfeta}

\newcommand{\BgO}{\bOmega}

\begin{document}

\title{Scalar perturbations during multiple-field slow-roll inflation}

\author{S Groot Nibbelink\dag\ and B J W van Tent\ddag}

\address{\dag Physikalisches Institut, Universit\"at Bonn,
Nu\ss allee 12, D-53115 Bonn, Germany}

\address{\ddag Spinoza Institute/Institute for Theoretical Physics, 
Utrecht University, P.O.\ Box 80195, 3508 TD Utrecht, The Netherlands}

{\vspace*{5pt}\address{E-mail: {nibblink@th.physik.uni-bonn.de},\ 
{B.J.W.vanTent@phys.uu.nl}}}

\begin{abstract}
We calculate the scalar gravitational and matter perturbations in the context 
of slow-roll inflation with multiple scalar fields, that take values on a 
(curved) manifold, to first order in slow roll. For that purpose a basis for
these perturbations determined by the background dynamics is introduced and
multiple-field slow-roll functions are defined. To obtain analytic solutions to
first order, the scalar perturbation modes have to be treated in three
different regimes. Matching is performed by analytically identifying leading
order asymptotic expansions in different regions. Possible sources for 
multiple-field effects in the gravitational potential are the particular 
solution caused by the coupling to the field perturbation perpendicular to the 
field velocity, and the rotation of the basis. The former can contribute even
to leading order if the corresponding multiple-field slow-roll function is
sizable during the last 60 e-folds. Making some simplifying assumptions, the
evolution of adiabatic and isocurvature perturbations after inflation is
discussed. The analytical results are illustrated and checked numerically with
the example of a quadratic potential. 
\end{abstract}

\pacno{98.80.Cq}



\section{Introduction}

As has been known for a long time, inflation  \cite{Guth,boekLinde} offers a
mechanism for the production of density perturbations,  which are supposed to
be the seeds for the formation of large scale structures in the universe.  This
mechanism is the magnification of microscopic quantum fluctuations in the
scalar fields present during the inflationary epoch into macroscopic matter and
metric perturbations. Also, since a part of the primordial spectrum of density
perturbations is observed in the cosmic microwave background radiation (CMBR),
this mechanism offers one of the most important ways of checking and
constraining possible models of inflation, see e.g.\ \cite{Kinneyetal},
especially when combined with large scale structure data \cite{Hannestadetal}.

The theory of the production of density perturbations in inflation has been 
studied for a long time. First in the case of a single real scalar field (see
e.g.\ \cite{Bardeenetal,Mukhanovetal,LiddleLyth,StewartLyth,MartinSchwarz,
StewartGong,Schwarzetal}), but later also for multiple fields, for example 
in the following papers. Pioneering work was done in 
\cite{KodamaSasaki,Starobinsky}. Using gauge invariant variables the authors of
\cite{PolarskiStarobinsky,GarciaWands,Garciaetal,NambuTaruya,Gordonetal} 
treated two-field inflation models.
The fluid flow approach was extended to multiple fields in  \cite{LythRiotto},
while a more geometrical approach was used in
\cite{SasakiStewart,NakamuraStewart}; both methods assumed several
slow-roll-like conditions on the potential. Using slow-roll approximations for
both the background and the perturbation equations the authors of
\cite{MukhanovSteinhardt,PolarskiStar,Langlois} were able to find expressions 
for the metric perturbations in multiple-field inflation.
The authors of \cite{HwangNoh1,HwangNoh} paid special attention to gauge issues
in their discussion of multiple-field perturbations.
The case of perturbations in generalized gravity theories was
studied in \cite{StarYokoyama,HwangNohgrav,StarTsuYok}.

There are two important reasons for considering inflation with multiple scalar
fields. To realize sufficient inflation before a graceful exit from the
inflationary era and produce the observed density perturbation spectrum in a
model without very unnatural values of the parameters and initial conditions, 
one is naturally led to the introduction of additional fields.
This is the motivation for hybrid inflation models \cite{Linde} (related
models can be found in \cite{LythRiotto}). The other reason is that many 
theories beyond the standard model of particle physics, like grand unification, 
supersymmetry or effective supergravity from string theory, contain a lot of 
scalar fields. Ultimately one would hope to be able to identify those fields 
that can act as inflatons. 
In addition such string-inspired supersymmetric models naturally have 
non-minimal kinetic terms due to a K\"ahler potential that is not of the 
minimal form $G = \bz z$ for a complex scalar field $z$ (perhaps a modulus 
of string theory) contained in a chiral multiplet. 
(Since a complex field can always be written as two real fields, it is 
sufficient to consider only real scalar fields in our paper.)

To be able to investigate the possibility that such models can produce 
(sufficient) inflation and density perturbations in accordance with 
present and future observations, it is necessary to have a general treatment 
available that can handle an arbitrary number of scalar fields with an 
arbitrary (K\"ahler) field metric and a generic potential. Most of the previous
literature on multiple-field inflationary density perturbations is limited with
respect to these aspects, usually by considering only two fields and minimal
kinetic terms. An exception to this are the papers 
\cite{SasakiStewart,NakamuraStewart}, but these still left space for
improvement, most importantly regarding the treatment of slow roll, the 
rotation of background fields, the transition region, and the analysis of the
particular solution for the gravitational potential caused by the coupling to
multiple fields.
In this paper we provide a general treatment by computing the scalar 
gravitational and matter perturbations to first order in slow roll during 
inflation with multiple real scalar fields that may have non-minimal kinetic 
terms. Which of these fields acts as inflaton during which part of the
inflationary period is determined automatically in our formalism and does not
have to be specified beforehand. 
The treatment has three important ingredients: a geometrical setup, generalized 
slow-roll functions, and a thorough discussion of the period when the scales of
the perturbation modes are of the same order of magnitude as the Hubble scale. 
Let us elaborate on these points. 

First of all, the scalar fields and their perturbations have to 
be described in a geometrical way to make certain that their physical 
description is independent of the coordinate system that parameterizes 
the field manifold. To describe the evolution and 
quantization of the perturbations it is essential to introduce a
background-field induced basis (of the tangent bundle), such that the 
components of the perturbations in this basis are canonically normalized 
so that only physical degrees of freedom are quantized. This basis is also 
used to distinguish effectively single-field and truly multiple-field effects 
and to identify adiabatic and entropy perturbations during inflation.

We generalize the slow-roll parameters for a single background
field to multiple scalar fields in a systematic way. 
They are defined independently of the 
coordinates used on the scalar manifold. In addition, our definition of these 
variables is such that they apply to any background time variable (e.g.\ 
comoving or conformal time). In our paper we need this generality 
to rigorously define the notion of slow roll applied to the perturbations. 
Our slow-roll functions are defined independently of whether slow roll is 
valid, and can be viewed as short-hand notation to denote derivatives of the 
Hubble parameter and the field velocity. As such they 
can be identified in all kinds of exact equations, making it clear what 
the behaviour of these equations is if the system is in the slow-roll regime. 
The relevance of the slow-roll function $\tget^\perp$ should be stressed at 
this point, as it is the measure of true multiple-field effects. Indeed, by 
definition $\tget^\perp$ vanishes in the case of a single scalar field model. 
The multiple-field character of inflation is only apparent when $\tget^\perp$ 
is non-negligible. In particular, it determines the size of the multiple-field 
contributions to the adiabatic perturbation and the mixing between adiabatic 
and isocurvature perturbations.

During inflation there is a relatively sharp transition in the behaviour of a
fluctuation when the corresponding wavelength becomes larger than the Hubble
radius (`passes through the horizon'); this moment identifies a certain 
scale $k$. For observationally interesting scales from the point of view of
inflation (those that reentered the horizon only after the time of recombination
when the CMBR was formed) this happened approximately 60 e-folds before the end 
of inflation \cite{LiddleLyth}.
To find results for the perturbations valid to first order in slow roll this
transition has to be treated very carefully. We determine the order of various 
contributions, in particular multiple-field rotation effects, to see if they can
be neglected to first order or should be taken into account. Analytic
properties of the solutions in the different regions are used to relate the
normalization factor at the end of inflation to the initial conditions at a much
earlier time before the transition.

The main subject of this paper is the treatment of perturbations during
inflation, but we also provide results for adiabatic and isocurvature
perturbations and the correlations between them at the time of recombination 
when the CMBR was formed. Although adiabatic perturbations are always present, 
this is not true for isocurvature perturbations. Whether the latter survive till
recombination at all depends on additional physical assumptions.
Here we  follow \cite{Langlois} regarding assumptions about the decay of
the scalar  fields and the evolution after inflation, which are such that
isocurvature perturbations do survive in principle. We also neglect the effects
of preheating. A more thorough treatment of these aspects will be the subject
of another paper.

Apart from this introduction the paper is structured as follows. 
In section~\ref{backgrsec} the background 
with multiple scalar fields is described using geometrical concepts which 
are explained in \ref{geometry}. An orthonormal basis induced by
the dynamics of the background fields is also introduced here. 
Section~\ref{secSR} then describes the multiple-field slow-roll formalism.

Section~\ref{multipert} is devoted to the perturbations in multiple-field
inflation and is the main part of this paper. In section~\ref{eqmotpert} 
the equations of motion for the scalar gravitational and matter perturbations
are derived, and the choice of perturbation variables is discussed.
The next section~\ref{quantization} focuses on the quantization of the 
dynamical scalar perturbations. 
After discussing the outlines of the calculation in section~\ref{pertsolsetup},
and introducing the concept of slow roll on the perturbations in
section~\ref{slowrollpert}, solving the equations to find expressions for the
perturbations, in particular the gravitational potential, in terms of 
background quantities only is done in section~\ref{pertsolcalc}. 
Section~\ref{afterinfl} then deals with the perturbations after inflation and
gives expressions for the vacuum correlators of the gravitational potential
at the time of recombination.

In section~\ref{flatmasses} the example of a quadratic potential with multiple
scalar fields is discussed, not only to illustrate the theory of 
section~\ref{multipert}, but also as a numerical check of our analytical 
results. Analytical expressions for this example are derived in 
section~\ref{analytics}, while section~\ref{numerics} gives numerical results.
The results of this paper are summarized and discussed in 
section~\ref{conclusions}.

\section{Slow-roll background in multiple-field inflation}
\labl{multi}

\subsection{Equations of motion for the background}
\labl{backgrsec}

The background of the universe is described by the flat Robertson-Walker 
metric in terms of a general time variable $\gt$: 
\equ{
\d s^2 = - b^2 \d \gt^2 + a^2\d \vc x^2
\labl{backmetric}
}
with $a(\gt)$ the spatial scale factor. The temporal scale factor $b$ is
defined by the specific choice of time variable: for comoving time $t$ 
and conformal time $\get$ it is given by $b = 1$ and $b = a$, respectively,
leading to the relation $\d t = a \d\eta$.
The main reason for setting up the background equations in terms of a general
time variable is that we need those in our discussion of slow roll on the
perturbations in section~\ref{slowrollpert}. In addition,
since different equations are best solved using different time variables, it is
convenient to set up the formalism for a general time variable, to avoid
repetitions of almost identical equations. Moreover, with this
approach various definitions and conclusions are manifestly independent of the
choice of time coordinate, in particular those related to the slow-roll 
approximation.
A derivative with respect to the general time variable $\gt$ is denoted 
by $^; \equiv \der_\gt$, one with respect to comoving time by 
$\dot{} \equiv \der_t$, and one with respect to conformal time by 
$^\prime \equiv \der_\get$.
Hubble parameters $H_a \equiv \der_\gt a/ a$ and $H_b \equiv \der_\gt b/ b$ 
are associated with the scale factors $a$ and $b$. For $H_a$ in terms of 
comoving and conformal time we define the conventional symbols:
$H = \dot a/ a$ and $\cH = a'/ a = a H$. 

For the matter part of the universe we consider an arbitrary number of real 
scalar fields that are the components of a vector $\Bgf$ and the coordinates 
on a possibly non-trivial field manifold $\cM$ with metric $\mx{G}$. 
The Lagrangean for the scalar field theory with a potential $V$ on 
this manifold in a general spacetime that is quadratic in the derivatives 
can be written as
\equ{
\fl
\cL_\cM = 
\sqrt{-g} \lh - \half \der^\gm \Bgf \cdot \der_\gm\Bgf - V(\Bgf) \rh
=
\sqrt{-g} \lh - \half g^{\mu\nu} \der_\mu \Bgf^T \mx G \der_\nu \Bgf
- V(\Bgf) \rh,
\labl{Lagrangean}
}
with $g$ the determinant of $g_{\mu\nu}$.
Notice that the kinetic term contains both the inverse spacetime metric 
$g^{\mu\nu}$ and the field metric $\mx{G}$.
Definitions of various geometrical concepts like the inner product 
$\vc{A}\cdot\vc{B}=\vc{A}^\dag\vc{B}=\vc{A}^T\mx{G}\vc{B}$
and the derivatives $\cD_\gm$ (with respect to 
spacetime) and $\Bnabla$ (with respect to the fields), that are
covariant with respect to the geometry of the manifold $\cM$, can be found 
in \ref{geometry}.

This geometrical description of the kinetic part 
of the Lagrangean is motivated by high-energy theoretical 
models where non-minimal kinetic terms appear naturally. In particular, in 
supergravity models \cite{Cremmer:1983en,Nilles:1984ge}, 
in which renormalizability is no longer an issue, 
supersymmetry forces the scalar Lagrangean to take the form
\equ{
\cL = - G_{\bga\ga} \der_\gm \bz^\uga \der^\gm z^\ga - V, 
\qquad\qquad
V = e^{-G} \Bigl( G_\ga G^{\ga \uga} G_{\uga} -3 \Bigr).
}
Here the subscripts $_\ga$ and $_\uga$ denote differentiation 
with respect to the complex coordinates $z^\ga$ and their conjugates 
$\bz^\uga$, respectively. These coordinates parameterize a 
so-called K\"ahler manifold, which has the property that the 
metric $G_{\uga \ga}$ (with $G^{\ga \uga}$ its inverse) 
can be determined from the K\"ahler potential $G$ as its second 
mixed derivative. This K\"ahler potential $G = K + \ln |W|^2$ 
consists of a K\"ahler potential $K(z,\bz)$ of the complex 
manifold and a holomorphic superpotential $W(z)$.
Models of (heterotic) string theory obtained by compactification 
lead to supergravity theories with the structure of the scalar field 
theory described above, see \cite{Witten:1985xb}. The resulting K\"ahler 
potential $K$ has typical no-scale supergravity features 
\cite{Lahanas:1987uc} due to the strong requirement of modular 
invariance \cite{Ferrara:1989bc}. An introduction to string 
phenomenology can be found in \cite{Quevedo:1996sv}.

The equations of motion for the scalars are given by
\equ{
g^{\mu\nu} \lh \cD_\mu \gd^\gl_\nu - \gG^\gl_{\mu\nu} \rh \der_\gl \Bgf 
- \mx G\inv \Bnabla^T V = 0,
\labl{eqmotfull}
} 
and the Einstein equations read 
\equ{
\frac 1{\gk^2} G^\gm_{\; \nu} = T^\mu_{\;\nu} 
= \der^\mu \Bgf \cdot \der_\nu \Bgf
- \gd^\mu_\nu \lh \half \der^\gl \Bgf \cdot \der_\gl \Bgf + V \rh,
\labl{Tmunu2}
}
with $G^\gm_{\; \nu}$ the Einstein tensor and 
$\gk^2 \equiv 8\pi G = 8\pi/M_P^2$.
From these formulae \eqref{eqmotfull} and \eqref{Tmunu2} we obtain 
the background equation of motion for the scalar fields $\Bgf$, 
\equ{
\cD {\Bgf}^; + 3 H_a {\Bgf}^; 
+ b^2\, \mx G\inv \Bnabla^T V = 0,
\labl{eqmotback}
}
and the Friedmann equations 
\equ{
H_a^2 = \frac{1}{3} \gk^2 \lh \half |{\Bgf}^;|^2 + b^2 V \rh,
\qquad\qquad
\cD {H_a}  = - \frac{1}{2} \gk^2 |{\Bgf}^;|^2.
\labl{Friedmann}
}
Here we have introduced the ``slow-roll derivative'' $\cD$ which is 
defined as follows: on any quantity $A$ that does not have any $b$ dependence, 
$\cD (b^n A) = (\cD_\gt - n H_b) (b^n A)$. In particular this means that
$\cD \Bgf^; = (\cD_\gt - H_b) \Bgf^;$,  
$\cD^2 \Bgf^; = (\cD_\gt - 2 H_b) (\cD_\gt - H_b) \Bgf^;$,  
$\cD H_a = (\der_\gt - H_b) H_a$, etc. 
Notice that the slow-roll derivative equals the comoving time derivative $\cD_t$
if comoving time is used ($b = 1$), while with conformal time it reads 
$\cD = \cD_\get - n \cH$. 
The slow-roll derivative is a necessary ingredient to be able to write these
equations in terms of a general time variable. It has the important property
that when it is applied to quantities like field velocities or Hubble
parameters, only terms of one order higher in the slow-roll approximation are
obtained (hence its name), as we show in the next section.

We finish this section by introducing a prefered basis $\{\vc{e}_n\}$ on the
field manifold that is induced by the dynamics of the system. 
This basis was already introduced in our previous paper \cite{SGNBJWvT},
although not in the context of a general time variable. The treatment with the
angle of \cite{Gordonetal} is a special case of this basis in the limit of only
two fields.
The first unit  vector $\vc{e}_1$ is given by the
direction of the field velocity $\Bgf^;$. The second unit vector 
$\vc e_2$ points in the direction of that part of the field acceleration 
$\cD \Bgf^;$ that is perpendicular to the first unit vector $\vc e_1$.  
This Gram-Schmidt orthogonalization process can be extended to any $n$: 
the unit vector $\vc{e}_n$ points in the direction of 
\(
\Bgf^{(n)} \equiv \cD^{(n-1)} \Bgf^;
\)
that is perpendicular to the first $n-1$ unit vectors $\vc{e}_1,\ldots, 
\vc{e}_{n-1}$.
Using the projection operators $\mx{P}_n$, which project on the $\vc{e}_n$, and  
$\mx{P}_n^\perp$, which project on the subspace that is perpendicular to 
$\vc{e}_1, \ldots, \vc{e}_n$, the definitions of the unit vectors are given by
\equ{
\vc{e}_n =
\frac{ \mx{P}_{n-1}^\perp \Bgf^{(n)} }{ |\mx{P}_{n-1}^\perp \Bgf^{(n)}| },
\qquad\qquad 
\mx{P}_n = \vc{e}^{\;}_n \vc{e}_n^\dag, 
\qquad\qquad
\mx{P}_n^\perp = \Id - \sum_{q=1}^n \mx{P}_q,
\labl{basisconstr}
}
for all $n = 1, 2, \ldots$ and with the definition $\mx{P}_{0}^\perp \equiv \Id$.
Notice that the unit vectors $\vc{e}_n$ will in general depend on time.
However, because the slow-roll derivative $\cD$ was used in the 
definition of this basis, the definition does not depend on a specific choice 
of time variable.
By construction the vector $\Bgf^{(n)}$ can be expanded 
in these unit vectors as 
\equ{
\Bgf^{(n)} = 
\left( 
\mx{P}_1 + \ldots + \mx{P}_{n} 
\right) \Bgf^{(n)} = 
\sum_{p = 1}^n \gf^{(n)}_{~p} \vc{e}_p,
\qquad\qquad
\gf^{(n)}_{~p} = \vc{e}_p \cdot \Bgf^{(n)}.
\labl{constructBgfn}
}
In particular, we have that 
\(
\gf^{(n)}_{~n} = \vc{e}_n \cdot \Bgf^{(n)} = 
| \mx{P}_{n-1}^\perp \Bgf^{(n)} |.
\)
As the projection operators $\mx{P}_1$ and $\mx{P}_1^\perp$ turn out to 
be the most important in our discussions, we introduce the short-hand notation
\(
\mx{P}^\parallel = \mx{P}_1
\)
and 
\(
\mx{P}^\perp = \mx{P}_1^\perp = \Id - \mx{P}^\parallel.
\)
In terms of these two operators we can write a general vector and matrix as
\(
\vc{A} = \vc{A}^\parallel + \vc{A}^\perp
\)
and 
\(
\mx{M}  =  \mx{M}^{\parallel\parallel} + \mx{M}^{\parallel\perp} +  
\mx{M}^{\perp\parallel} + \mx{M}^{\perp\perp},
\)
with $\vc A^\parallel \equiv \mx P^\parallel \vc A$ and 
$\mx{M}^{\parallel\,\parallel} \equiv 
\mx{P}^\parallel \mx{M} \mx{P}^\parallel$, etc.  

\subsection{Slow roll}
\labl{secSR}

Slow-roll inflation is driven by a scalar field potential that is almost flat
and therefore acts as an effective cosmological constant. In the case of a
single scalar field, the notion of slow roll is  well-established (see e.g.\
\cite{LiddleLyth, LythRiotto, LiddleParsonsBarrow}). This concept can be
generalized to multiple scalar fields in a geometrical way using the unit
vectors introduced in the previous section.
The system consisting of \eqref{eqmotback} and \eqref{Friedmann} 
is said to be in the slow-roll regime if the comoving time derivatives satisfy
$|\cD_t \Dot{\Bgf}| \ll |3 H \Dot{\Bgf}|$ and $\half |\Dot{\Bgf}|^2 \ll V$. 
A more precise definition not depending on the use of 
comoving time is given below \eqref{eqmotbackiter}. 

We introduce the following functions for an arbitrary time variable $\gt$ (see
also \cite{SGNBJWvT}): 
\equ{
\tge(\Bgf) \equiv  
- \frac{\cD H_a}{H_a^2} ,
\qquad\qquad
\tilde{\Bget}^{(n)} (\Bgf) \equiv 
\frac {\cD^{n-1} {\Bgf}^;}{(H_a)^{n-1} | \Bgf^; | }. 
\labl{slowrollfun}
}
We often use the short-hand notation $\tilde{\Bget} = \tilde{\Bget}^{(2)}$ and 
$\tilde{\Bgx} = \tilde{\Bget}^{(3)}$. Both these vectors can be 
decomposed in components parallel ($\tget^\parallel, \tgx^\parallel$) 
and perpendicular ($\tget^\perp$) to the field velocity $\Bgf^;$:
\equ{
\fl
\tget^\parallel 
= \vc{e}_1 \cdot \tilde{\Bget} = 
\frac{\cD {\Bgf}^; \cdot {\Bgf}^;} {H_a |{\Bgf}^;|^2}, 
\qquad
\tget^\perp
= \vc{e}_2 \cdot \tilde{\Bget} = 
\frac{|(\cD {\Bgf}^;)^\perp|}{H_a |{\Bgf}^;|}, 
\qquad
\tgx^{\parallel}
= \vc{e}_1 \cdot \tilde{\Bgx} = 
 \frac{\cD^2 {\Bgf}^; \cdot {\Bgf}^;}
{H_a^2 |{\Bgf}^;|^2}. 
\labl{slowrollfuncomp}
}
(Even though $\tilde{\Bgx}$ in general has two directions perpendicular to 
$\vc{e}_1$ we only give $\gx^\parallel$ here since it is the only one 
that turns out to be relevant in the remainder of this work.)
The derivatives of the slow-roll functions can be computed from their
definitions and are given by:
\equ{
\fl
\tge^; = 2 H_a \tge ( \tge + \tget^\parallel ), 
\quad
(\tget^\parallel)^; = H_a [ \tgx^\parallel 
+ (\tget^\perp)^2 + \tge \tget^\parallel - (\tget^\parallel)^2 ],
\;
\cD \tilde{\Bget} = H_a [ \tilde{\Bgx} + 
( \tge - \tget^\parallel) \tilde{\Bget} ].
\labl{dereps}
}

In terms of the functions $\tge, \tilde{\Bget}$ the Friedmann equation
\eqref{Friedmann} and the background field equation \eqref{eqmotback}
read
\equ{
\fl
H_a = \frac{\gk}{\sqrt{3}} \, b\, \sqrt{V} \lh 1 - \frac{1}{3} \tge
\rh^{-1/2},
\labl{Friedmanniter}\\
\fl
{\Bgf}^; + \frac{2}{\sqrt{3}} \, 
\frac{1}{\gk} \,b\, \mx G\inv \Bnabla^T \sqrt{V}
= - \sqrt{\frac{2}{3}} \,b\,  \sqrt{V} \, \frac{\sqrt{\tge}}{1-\frac{1}{3}\tge}
\lh \frac{1}{3} \tilde{\Bget} + \frac{\frac{1}{3} \tge \, \vc{e}_1}
{1+\sqrt{1-\frac{1}{3}\tge}} \rh.
\labl{eqmotbackiter}
}
(Notice that for a positive potential $V$ the function $\tge < 3$, 
as can be seen from its definition.) We can now define precisely 
what is meant by slow roll as these two background equations 
are still exact. Slow roll is valid if $\tge$, 
$\sqrt{\tge}\, \tget^\parallel$ and $\sqrt{\tge}\, \tget^\perp$ are (much) 
smaller than unity. For this reason $\tge$, $\tget^\parallel$ and $\tget^\perp$
are called slow-roll functions. The function $\tilde{\Bgx}$  is called a second 
order slow-roll function because it involves two slow-roll derivatives, 
and it is assumed to  be of an order comparable to $\tge^2$, 
$\tge\tget^\parallel$, etc. If slow roll is valid, we can use expansions in 
powers of these slow-roll functions to estimate the relevance of various 
terms in a given expression. For example, to first order the 
Friedmann equation \eqref{Friedmanniter} is approximated by
replacing $(1-\tge/3)^{-1/2}$ by $(1+\tge/6)$.
The background field equation up to and including first order is given by 
\eqref{eqmotbackiter} with the right-hand side set to zero, as all 
those terms are order $3/2$ or higher. This last fact is the motivation for
defining results accurate to first order to have no corrections larger than
order $\tge^{3/2}$ when discussing the perturbations in
section~\ref{pertsolsetup}.
 
At the level of the solutions of these equations we make the following
definition. An approximate solution of an equation of motion is said to be 
accurate to first order in slow roll, if the relative difference between this 
solution and the exact one is of a smaller numerical order than the slow-roll
functions. This relative error depends in general on the size of the 
integration interval. Let us explain this with the following example.
From \eqref{dereps} we see that the time derivatives of the slow-roll functions
are second order quantities. Hence we can make the assumption that to first
order the slow-roll functions are constant. Switching to the number of e-folds
$N$, which is related as $\d N = H_a \d\gt$ to the time variable $\gt$, we can 
integrate \eqref{dereps} to find the variation of $\tge$ over an 
interval $[N_1,N_2]$:
\equ{
\gD \tge = \int_{N_1}^{N_2} \d N \, 2 \tge (\tge+\tget^\parallel)
= 2 \tge_0 ( \tge_0 + \tget^\parallel_0 ) (N_2 - N_1).
\labl{vareps}
}
Here the subscript $_0$ denotes some reference time in this interval where the
slow-roll functions are evaluated. Hence we see that if the interval
$(N_2-N_1)$ becomes larger than $1/(2(\tge_0+\tget^\parallel_0))$, $\gD\tge$
becomes larger than $\tge_0$ and the assumption of taking $\tge$ constant over
this interval to first order is certainly not valid anymore. (An example of the
real behaviour of the slow-roll functions can be found in
figure~\ref{fieldfig}b) in section~\ref{numerics}.) 
In the literature these effects are usually ignored and the solution of an
equation of motion valid to first order is (implicitly) assumed to be accurate
to first order as well. However, with that assumption the numerical error 
between slow-roll and exact solution can become very large depending on the size
of the interval of integration, which is the reason for our revised definition.

The slow-roll functions (\ref{slowrollfun}) are all defined as  functions of
covariant derivatives of the velocity $\Bgf^;$ and the Hubble parameter $H_a$.  
If the zeroth order slow-roll
approximation works well,  that is if the right-hand side of
\eqref{eqmotbackiter} can be neglected, as  well as the $\tge$ in
\eqref{Friedmanniter}, then we can use these two equations to eliminate
${\Bgf}^;$ and $H_a$ in favour of the potential $V$.  This is the way the
original single-field slow-roll parameters were defined. However, that
original definition had the disadvantage that the slow-roll  conditions
became consistency checks. While we can expand the exact equations in
powers of the slow-roll functions, that is impossible by construction with the
original slow-roll parameters in terms of the potential.\footnote{In the 
context of single-field inflation this was discussed in 
detail in \cite{LiddleParsonsBarrow}.}
In order to avoid confusion we compare the slow-roll functions we defined in
\eqref{slowrollfun} with the ones originally used in the single-field
case, $\ge$ and $\get$:
\equ{
\ge = \frac{1}{2 \gk^2} \frac{V_{,\gf}^2}{V^2} = \tge, 
\qquad\qquad
\get = \frac{1}{\gk^2} \frac{V_{,\gf\gf}}{V} = - \tget^\parallel + \tge,
\labl{convsrpar}
}
where the last equalities in both equations are only valid
to lowest order in the slow-roll approximation. 

For later use we introduce the matrix $Z$ by 
\equ{
(Z)_{mn} = - (Z^T)_{mn} =  \frac 1{H_a} 
\vc e_m^\dag \cD \vc e_n^{\,},
\labl{mtrxZ}
}
which shows a nice interplay between the unit vectors and the notion of slow
roll. The anti-symmetry of $Z$ follows because 
$(\vc{e}_m^\dag \vc{e}_n^{\,})^;=0$. To determine its components 
we observe that 
\equ{
\fl
\cD \vc e_{n+1}\cdot \vc e_{n - p} + 
 \vc e_{n+1}\cdot \cD \vc e_{n - p} = 0, 
\qquad
\cD \vc e_{n+1} \cdot \tilde{\Bget}^{(n)} + 
H_a\, \vc e_{n+1} \cdot \tilde{\Bget}^{(n+1)} = 0,
\labl{Deetc}
}
because $\vc e_{n+1}$ is perpendicular to $\vc e_{n-p}$ with $0 \leq p < n$
and to $\tilde{\Bget}^{(n)}$. From the construction of $\Bgf^{(n)}$ in
\eqref{constructBgfn} we see that $\cD\vc{e}_n$ can never get a component in a
direction higher than $\vc{e}_{n+1}$. Hence we deduce from the first equation
in \eqref{Deetc} that for $p \geq 1$, $\cD \vc e_{n+1}$ and $\vc e_{n - p}$ are 
perpendicular. Using this we see that of the first term of the second 
equation only the $\vc e_n$ direction is relevant, so that the only 
non-zero components of $Z$ read
\equ{
Z_{n\, n+1} = - Z_{n+1 \, n} = -
\frac {\vc e_{n+1}\cdot  \tilde{\Bget}^{(n+1)}}
{\vc e_{n} \cdot \tilde{\Bget}^{(n)}},
\labl{Zcomp}
}
which is first order in slow roll.

\section{Perturbations in multiple-field inflation}
\labl{multipert}

\subsection{Equations of motion for the perturbations}
\labl{eqmotpert}

This section describes the coupled system of gravity, encoded by
the metric $g_{\mu\nu}$, and multiple scalar field perturbations $\Bgd\Bgf$
during inflation. We separate both the scalar fields and the
metric into a homogeneous background part and an inhomogeneous perturbation,
which is assumed to be small. Since the observed fluctuations in the CMBR are
tiny, this assumption is well-motivated. Consequently one can linearize
all equations with respect to the perturbations. We define
\equ{
\eqalign{
\fl
\Bgf^{\mathrm{full}}(\eta,\vc{x}) = \Bgf(\eta) + \Bgd\Bgf(\eta,\vc{x}),
\labl{separation}\\
\fl
g_{\mu\nu}^{\mathrm{full}}(\eta,\vc{x}) 
= g_{\mu\nu}(\eta) + \gd g_{\mu\nu}(\eta,\vc{x}) =
a^2(\eta) \pmtrx{-1 & 0\\ 0 & \gd_{ij}}
- 2 a^2(\eta) \Phi(\eta,\vc{x}) \pmtrx{1 & 0\\ 0 & \gd_{ij}}.
}}
As is discussed in \cite{Mukhanovetal}, this metric is obtained by applying the
so-called longitudinal gauge to the flat Robertson-Walker metric in the case
when only scalar metric perturbations and a scalar matter theory are
considered. In this gauge all formulae look the same as when the
gauge-invariant approach \cite{Bardeen, Mukhanovetal} is used. 
The gravitational (Newtonian) potential $\Phi(\eta,\vc{x})$
describes the scalar metric perturbations.

The equation of motion for the perturbations of the metric is 
obtained by linearizing and combining the $(00)$ and $(ii)$ components of 
the Einstein equations \eqref{Tmunu2}: 
\equ{
\Phi'' + 6 \cH \Phi' + 2 (\cH' + 2\cH^2) \Phi - \gD \Phi
= - \gk^2 a^2 (\Bnabla V \, \Bgd\Bgf),
\labl{eqPhi2}
}
where the spatial Laplacean is given by $\gD = \sum_i \der_i^2$,  
while the integrated $(0i)$ component of the Einstein equations leads to the
constraint equation
\equ{
\Phi' + \cH \Phi = \half \gk^2 \Bgf' \cdot \Bgd\Bgf 
= \half \gk^2 |\Bgf'| \gd\gf^\parallel.
\labl{Einstein0i2}
}
Here we have decomposed $\Bgd\Bgf = \gd\gf^\parallel \, \vc{e}_1 +
\Bgd\Bgf^\perp$. (A similar decomposition in the case of two-field
inflation was also discussed in \cite{Gordonetal}.)  
In addition we have the equation of motion for the scalar field perturbations,
\equ{
\lh \cD_\eta^2 + 2 \cH \cD_\eta - \gD 
+ a^2 \tilde{\mx{M}}^2(\Bgf) \rh \Bgd\Bgf
=  4 \Phi' \Bgf' - 2 a^2 \Phi\, \mx G\inv \Bnabla^T V,
\labl{eqmotpert2}
}
where we have introduced the (effective) mass-matrices
\equ{
\tilde{\mx M}^2 \equiv \mx M^2 - \mx R (\Dot{\Bgf},\Dot{\Bgf}),
\qquad\qquad
\mx M^2 \equiv \mx G\inv \Bnabla^T \Bnabla V,
\labl{massmatrix}
}
with $\mx{R}$ the field curvature as defined in the appendix.
This system of perturbation equations must be solved in the background 
determined by the scalar fields \eqref{eqmotback} and the Friedmann 
equations \eqref{Friedmann}. 
Using  the integrated $(0i)$ component of the Einstein equations 
\eqref{Einstein0i2} together with the background equation of motion for the
scalar fields \eqref{eqmotback},
the right-hand side of equation \eqref{eqPhi2} for $\gF$ can be rewritten as
\equ{
\fl
- \gk^2 a^2 (\Bnabla V \, \Bgd\Bgf)  = 
2 (\Phi' + \cH \Phi) \lh \frac{1}{|\Bgf'|} (\cD_\eta \Bgf')
\cdot \vc{e}_1 + 2 \cH \rh
+ \gk^2 (\cD_\eta \Bgf') \cdot \Bgd\Bgf^\perp,
}
where we used the definition of the projection operators.
Inserting this expression in \eqref{eqPhi2} and realizing that 
\(
|\Bgf'|' \, |\Bgf'| = (\cD_\eta \Bgf') \cdot \Bgf',
\)
we get
\equ{
\fl
\Phi'' + 2 \lh \cH - \frac{|\Bgf'|'}{|\Bgf'|} \rh \Phi'
+ 2 \lh \cH' - \cH \frac{|\Bgf'|'}{|\Bgf'|} \rh \Phi - \gD \Phi
= \gk^2 (\cD_\eta \Bgf') \cdot \Bgd\Bgf^\perp.
\labl{eqPhi2a}
}
In the single-field case the right-hand side is zero because $\Bgd\Bgf^\perp$
then vanishes by construction.

The system of perturbations \eqref{eqPhi2a}, \eqref{Einstein0i2} and 
\eqref{eqmotpert2} is quite complicated. To make the physical content 
more transparent, we introduce new variables $u$ and $\vc q$ (linearly 
related to $\gF$ and $\Bgd\Bgf$, respectively),
\equ{
u \equiv \frac{a}{\gk^2 |\Bgf'|} \,  \Phi
= \frac{\Phi}{\gk \sqrt{2} \, H \sqrt{\tge}},
\qquad\qquad
\vc{q} \equiv a \lh \Bgd\Bgf + \frac{\Phi}{\cH} \, \Bgf' \rh,
\labl{newvars}
}
which satisfy the following two requirements: 
\enums{
\item The equations of motion for both $u$ and $\vc q$ do not 
contain first order conformal time derivatives; 
\item The equation of motion for $\vc q$ is homogeneous and $\vc q$ is gauge 
invariant.
}
The first requirement makes a direct comparison between the size of the 
Fourier mode $\vc k^2 = k^2$ and other physical background quantities 
in the equation of motion possible. In section \ref{pertsolsetup} we make use
of this to distinguish between different regions for the behaviour of the
solutions.
The other requirement ensures that we can naively 
quantize $\vc q$ using the Lagrangean corresponding to the equation 
of motion for $\vc q$ in section \ref{quantization}. 
As $\vc q$ is gauge invariant and linearly related to $\Bgd\Bgf$, 
apart from the shift proportional to $\gF \Bgf'$, no non-physical 
degrees of freedom are quantized. The single-field version of $\vc{q}$,
including its equation of motion and quantization, was first introduced by
Sasaki and Mukhanov \cite{Sasaki,Mukhanov}, which is why variables of this 
type are sometimes refered to as Sasaki-Mukhanov variables. The variable $u$ was
first introduced by Mukhanov \cite{Mukhanov1}, see also \cite{Mukhanovetal}.

To derive the equation of motion for $\vc q$ we need an auxiliary result. 
By differentiating the background field equation in terms of conformal 
time, $\cD_\eta \Bgf' + 2 \cH \Bgf' + a^2 \mx{G}\inv \Bnabla^T V = 0$, once 
more we obtain
\equ{
\cD_\eta^2 \Bgf' + 2(\cH' - 2 \cH^2) \Bgf' + a^2 \tilde{\mx{M}}^2 \Bgf' = 0,
}
where we used that $\cD_\eta (\mx{G}\inv \Bnabla^T V) = \mx{M}^2 \Bgf' =
\tilde{\mx{M}}^2 \Bgf'$ (because of the anti-symmetry properties of the 
curvature tensor $\mx R(\Dot\Bgf, \Dot\Bgf) \Bgf' = 0$).
The equation for $\vc{q}$ is then obtained from the equation of motion
\eqref{eqmotpert2} for $\Bgd\Bgf$ and \eqref{eqPhi2}, \eqref{Einstein0i2} 
for $\Phi$, using the projectors \eqref{basisconstr} and slow-roll functions 
\eqref{slowrollfun}. Combining this with the derivatives of $\cH$ from 
\eqref{Friedmann} and of the slow-roll functions \eqref{dereps}, 
we finally obtain the homogeneous equation for the spatial Fourier mode 
$\vc k$ of $\vc q$:  
\equ{
\fl
\cD_\eta^2 \vc{q}_{\vc k} + ( k^2 + \cH^2 {\BgO}) 
\vc{q}_{\vc k} = 0,
\qquad\qquad
L = \half \cD_\get \vc q_{\vc k}^\dag \cD_\get \vc q_{\vc k} 
- \half \vc q_{\vc k}^\dag ( k^2 + \cH^2 \BgO )\vc q_{\vc k}.
\labl{eqq}
}
Here $L$ is the associated Lagrangean and 
\equ{
{\BgO} \equiv \frac{1}{H^2} \tilde{\mx M}^2
- (2 - \tge) \Id 
-  2 \tge \Bigl( ( 3 + \tge) \mx P^\parallel 
+ \vc e_1 \tilde{\Bget}^\dag + \tilde{\Bget} \vc e_1^\dag \Bigr).
\labl{masseff}
} 
The $(n1)$ components of $\BgO$ can be expressed completely in terms of
slow-roll functions using
\equ{
\frac 1{H^2} \tilde{\mx M}^2 \vc e_1 = 
\frac 1{H^2} {\mx M}^2 \vc e_1 = 
3\, \tge\, \vc e_1 - 3 \,\tilde{\Bget} - \tilde{\Bgx}.
\labl{Me1}
}
The other components can in general not be expressed in terms of the
slow-roll functions introduced in the previous subsection. 

To derive the equation of motion for $u$ it is convenient to introduce the 
quantity $\gth$,  
\equ{
\gth \equiv \frac{\cH}{a |\Bgf'|} = \frac{\gk}{\sqrt{2}}\frac{1}{a\sqrt{\tge}}
\labl{gthdblgth}
\\
\fl
\Rightarrow\qquad 
\frac{\gth'}{\gth} = - \cH \lh 1 + \tge + \tget^\parallel \rh,
\qquad
\frac{\gth''}{\gth} = \cH^2 \lh 2\tge + \tget^\parallel 
+ 2 (\tget^\parallel)^2 - (\tget^\perp)^2 - \tgx^\parallel \rh,
\non
}
where we also gave the resulting expressions for its derivatives, and observe 
that the following relations hold for the slow-roll functions:
\equ{
\fl
\cH' = \cH^2 ( 1 - \tge),
\qquad\qquad
\frac{|\Bgf'|'}{|\Bgf'|} = \cH( 1 + \tget^\parallel),
\qquad\qquad 
(\cD_\eta \Bgf')^\perp = \frac{\sqrt{2}}{\gk} \, \cH^2 \sqrt{\tge} \, 
\tilde{\Bget}^\perp.
\labl{slowrollrel}
}
By substituting the definitions of $u$ and $\vc q$ in \eqref{eqPhi2a}, where we
first rewrite the relation between $\gF$ and $u$ as $\gF = \gk \sqrt{2} \cH
\sqrt{\tge} \, u /a$, and using the above expressions and 
the derivatives of the slow-roll functions given in \eqref{dereps}, we obtain  
\equ{
u_{\vc k}'' + \lh k^2  - \frac{\gth''}{\gth}\rh u_{\vc k}
=  \cH \tget^\perp \vc e_2 \cdot \vc q_{\vc k}.
\labl{equ}
}
Notice that all these equations are still exact, no slow-roll approximations
have been made.
From this one can draw the conclusion that at the level of the
equations the redefined gravitational potential $u$ decouples from the 
perpendicular components of the field perturbation $\vc q^\perp$ to leading 
order, but at first order  mixing between these perturbations appears. 

The equations of motion \eqref{eqq} and \eqref{equ} show that the different
spatial Fourier modes of both $\vc q$ and $u$ decouple. From now on we only 
consider one generic mode $\vc k$, so that we can drop the subscripts 
${}_{\vc k}$. Rewriting equation \eqref{Einstein0i2} in terms of the components 
$q_{n} \equiv \vc e_n \cdot \vc q$ of $\vc q$  and differentiating 
it once gives
\equ{
u'  - \frac {\gth'}{\gth} u = \half q_{1}
\qquad\Ra\qquad
u'' - \frac {\gth''}{\gth} u = 
\half \Bigl( q_{1}' + \frac{\gth'}{\gth} q_{1} 
\Bigr),
\labl{u1der}
}
where $\gth$ and its derivatives are given in \eqref{gthdblgth}.
This equation for $u''$ can be combined with the equation of 
motion \eqref{equ} for $u$ to give
\equ{
k^2  u =  \cH \tget^\perp q_{2} 
-\half \Bigl( 
q_{1}' + \frac{\gth'}{\gth} q_{1} 
\Bigr).
\labl{uindgv}
}
After $\vc{q}$ has been quantized, this expression can be used to relate 
it to $u$. (Although this
relation could in principle be used to compute $u$ at the end of inflation, 
its numerical implementation can be rather awkward because of cancellation of 
large numbers. In numerical situations it turns out to be more convenient to 
determine $u$ from its own equation of motion and only use \eqref{uindgv} to 
find the correct quantization and initial conditions.)

\subsection{Quantization of the perturbations}
\labl{quantization}

We start with the Lagrangean \eqref{eqq} in terms of the basis 
$\{ \vc e_n \}$:
\equ{
L = \half ( q' + \cH Z q)^T (q' + \cH Z q) - \half q^T (k^2 + \cH^2 \gO)q,
\labl{quantLagr}
}
where we employ the notation 
\(
(\gO)_{mn} = \vc e_m^\dag \BgO \vc e_n^{\,}
\)
and the matrix $Z$ is given in \eqref{mtrxZ}.
Notice that this Lagrangean has the standard canonical normalization of 
$\half (q')^T q'$, independent of the field metric $\mx{G}$, as can be 
derived from the original Lagrangean \eqref{Lagrangean}.
We maintain the vectorial structure of this multiple-field system
and repress the indices $n, m$ as much as possible, which means for example 
that the non-bold $q$ in this equation is a vector (in the basis
$\{\vc{e}_n\}$). 
From the canonical momenta $\gp=\der L/\der {q'}^T$ we find the Hamiltonian 
$H = \gp^T q' - L$ and the Hamilton equations: 
\equ{
H = \half (\gp - \cH Z q)^T (\gp - \cH Z q)
+ \half q^T \Bigl(k^2 + \cH^2 (\gO + Z^2) \Bigr) q; \non\\
q' = \pp[H] { \gp^T } = \gp - \cH Z q,
\qquad 
\gp' = - \pp[H] { q^T } = - (k^2 + \cH^2 \gO) q - \cH Z \gp.
\labl{HamEQ}
}

In order to avoid writing indices when considering commutation 
relations we use vectors $\ga, \gb$ with components 
$\ga_m$, $\gb_m$ in the $\vc e_m$ basis that are independent of $q$ and $\gp$. 
The canonical commutation relations can then be represented as
\equ{
[ \ga^T \hat q, \gb^T \hat q] = [ \ga^T \hat \gp, \gb^T \hat \gp] = 0,
\qquad\qquad
[ \ga^T \hat q, \gb^T \hat \gp] = i \ga^T \gb.
}
Using the Hamilton equations it can be checked that this quantization 
procedure is indeed time independent. 
Let $Q$ and $\gP$ be complex matrix valued solutions of the Hamilton equations,
such that $q= Q a^*_0 + \mbox{c.c.},~ \gp = \gP a_0^* + \mbox{c.c.}$
is a solution of \eqref{HamEQ} for any constant complex vector $a_0$.
Here $\mbox{c.c.}$ denotes the complex conjugate.
The Hamilton equations for $Q$ and $\gP$ can be combined to give a second order
differential equation for $Q$. To remove the first order time derivative from 
this equation, we define $Q(\get) = R(\get) \tQ(\get)$ 
with $R$ chosen such that the matrix functions $R$ and $\tQ$ satisfy
\equ{
\fl
R' + \cH Z R = 0, 
\qquad\qquad
\tQ''  + ( k^2  + \cH^2  \tilde{\gO} ) \tQ = 0, 
\qquad\mbox{with}\quad 
\tilde{\gO} = R\inv \gO R.
\labl{eqtq}
}
The matrix $\gP$ is then given by $\gP = Q' + \cH Z Q = R \tQ'$.
We take $R(\get_i) = \Id$ as initial condition, since 
the initial condition of $Q$ can be absorbed in that of $\tQ$.    
The equation of motion for $R$ implies that $R^T R$ and $\ln \det R$ 
are constant because $Z$ is anti-symmetric and consequently 
traceless. Taking into account its initial condition, it then follows 
that $R$ represents a rotation.

It now follows that $\hat q$ 
and $\hat \gp$ can be expanded in terms of constant creation ($\hat a^\dag$)
and annihilation ($\hat a$) operator vectors:
\equ{
\hat q = Q \hat a^\dag + Q^* \hat a 
= R \tQ \hat{a}^\dag + R \tQ^* \hat{a}, 
\qquad\qquad
\hat \gp = \gP \hat a^\dag + \gP^* \hat a.
\labl{defqwithR}
}
The creation and annihilation operators satisfy 
\equ{
[ \ga^T \hat a, \gb^T \hat a] = [ \ga^T \hat a^\dag, \gb^T \hat a^\dag] = 0,
\qquad\qquad
[ \ga^T \hat a, \gb^T \hat a^\dag] = \ga^T \gb.
}
This is consistent with the commutation relations for $q$ and $\gp$ given 
above, provided that the matrix functions $Q$ and $\gP$ satisfy
\equ{
Q^* \, Q^T - Q \, {Q^*}^T = \gP^*\, \gP^T - \gP \, {\gP^*}^T = 0,  
\qquad
Q^*\,  \gP^T - Q \, {\gP^*}^T = i \Id.
\labl{genWronskian}
}
These relations hold for all time, as can be checked explicitly by using the 
equations of motion for $Q$ and $\gP$ to show that they are time 
independent, provided that they hold at some given time. 

We assume that the initial state is the vacuum $|0\rangle$ defined by 
$\hat a | 0 \rangle = 0 $ and that there is no initial particle production. 
This implies that the Hamiltonian initially does 
not contain any terms with $\hat a \hat a$ and $\hat a^\dag \hat a^\dag$,
which leads to the condition
\equ{
( \gP - \cH Z Q )^T ( \gP - \cH Z Q )
+ Q^T \Bigl( k^2 + \cH^2( \gO - Z^T Z ) \Bigr) Q = 0.
\labl{no_prod}
}
The solution of the equations \eqref{genWronskian} and 
\eqref{no_prod} can be parametrized by a unitary 
matrix $U$ at the beginning of inflation, when the limit that 
$k^2$ is much bigger than any other scale is applicable:
\equ{
Q_i = \frac{1}{\sqrt{2 k}} \, U, 
\qquad\qquad
\gP_i = \frac {i \sqrt{k}}{\sqrt 2} \, U.
\labl{initQgP}
}

We denote expectation values with respect to the vacuum state $| 0 \rangle $ by 
$\langle \ldots \rangle $. Let $\ga, \gb$ be two vectors. Then for the 
expectation value of $(\ga^T Q U \hat a^\dag + \ga^{*T} Q^* U^* \hat a)^2$, 
with $U$ a unitary matrix, we obtain
\equ{
\langle (\ga^T Q U \hat a^\dag + \ga^{*T} Q^* U^* \hat a )^2 \rangle 
= \ga^{*T} Q^* U^* U^T Q^T \ga
= \ga^{*T} Q^* Q^T \ga. 
\labl{quadrexp}
}
So a unitary matrix in front of the $\hat a^\dag$ will drop out in the 
computation of this correlator. This is 
even true if another state than the vacuum is used to compute the correlator.
In particular this means that the correlator of the gravitational potential will
not depend on the unitary matrix $U$ in \eqref{initQgP}. To draw this conclusion
we use relation \eqref{uindgv} between $u$ and $\vc{q}$ and the fact that $Q$ 
satisfies a linear homogeneous equation of motion.
We also see that as long as $Q$ is simply oscillating and hence itself unitary
(apart from a normalization factor), its evolution will be irrelevant for the
computation of the correlator.

We finish this section with some brief remarks on the assumption of taking the
vacuum state to compute the correlator. 
The vacuum state $| 0 \rangle$ at the beginning  of
inflation seems a reasonable assumption for the calculation of the density
perturbations that we can observe in the CMBR today. Even though perturbations
in the  CMBR have long wavelengths now, they had very short wavelengths before
they crossed the Hubble radius during inflation. Therefore,  their scale $k$ at
the beginning of inflation at $t_i$ is much larger  than the Planck scale. It
seems a reasonable assumption that modes with momenta very much larger than 
the Planck scale are not excited at $t_i$, so that for these modes the vacuum
state is a good assumption.\footnote{There could be a problem with this
approach,  because our knowledge of physics beyond the Planck scale is
extremely poor.  In particular, the dispersion relation $\go(\vc{k})=k$ that we
used implicitly might not be valid for large $k$: there might be a cut-off for
large momenta.  For a discussion of this trans-Planckian problem and possible
cosmological consequences see \cite{MartinBrandenberger,Niemeyer}.}
This assumption can be tested by taking other states than the vacuum state. 
For instance one can try a thermal state with a temperature of the Planck 
scale. Typically one finds that if there were a few e-folds of inflation 
before the now observable scales crossed the Hubble scale, corrections are 
negligible. 
For a more detailed discussion on observable effects of non-vacuum initial 
states we refer to \cite{Lesgourguesetal,Martinetal}.

\subsection{Solutions of the perturbation equations to first order: setup}
\labl{pertsolsetup}

To derive analytical expressions for the gravitational potential and field
perturbations valid to first order in slow roll, we have to determine the 
evolution of the modified Newtonian potential $u$ and quantized variables 
$\vc q$, described by the equations \eqref{equ}, \eqref{eqq} and \eqref{u1der}, 
analytically and accurately up to first order during inflation. In this 
section we explain the physical ideas that go into that computation.
Section~\ref{slowrollpert} introduces the concept of slow roll on the
perturbations, which is useful in part of the calculations that are presented 
in section~\ref{pertsolcalc}.
The treatment here has been partly inspired by the discussion of the transition
region in \cite{MartinSchwarz}.

Since $\cH$ grows rapidly, while $k$ is constant for a given mode, the 
solutions of \eqref{equ} and \eqref{eqq} change dramatically around the time 
$\get_\cH$ when a scale crosses the Hubble scale. This time is defined by the
relation 
\equ{
\cH(\get_\cH) = k.
\labl{defetaH}
}
Notice that this means that $\eta_\cH$ depends on $k$.
Hence there are three regions 
of interest, which are denoted by their conventional names and treated in the 
following way: 
\items{
\item {\bf sub-horizon ($\cH \ll k$)}: 
This region is irrelevant for the computation of the correlators 
at the end of inflation (see \eqref{quadrexp}), since solving \eqref{eqtq} with 
the $\cH^2 \tgO$ term neglected with respect to the $k^2$ term we find 
\equ{
Q(\get) =  \frac 1{\sqrt{2k}}\, R(\get) \, e^{i k (\get-\get_i)} U
\qquad\Rightarrow\qquad
Q^*(\get) Q^T(\get) = \frac{1}{2k} \, \Id.
}
(Here the normalization is fixed by the initial condition  \eqref{initQgP}.)
The end of the sub-horizon period $\get_-$ is therefore defined as the 
moment when this does not hold any more to first order, leading to the
definition $\cH^2(\eta_-) = \tge^{3/2} k^2$.\footnote{The value $3/2$ is chosen 
here because that is the same order to which the slow-roll background field 
equation is valid, see \eqref{eqmotbackiter}, but the arguments are independent 
of which specific power (larger than one) is chosen.}
\item {\bf transition ($\cH \sim k$)}: 
We consider \eqref{eqtq} for $Q$, keeping all terms, but using that 
for a sufficiently small interval around $\eta_\cH$ 
the slow-roll functions can be taken to be constant to first order, which makes
it possible to obtain solutions for $Q$ valid to first order using Hankel 
functions. Since the effect of the sub-horizon region is irrelevant, we take 
the following initial conditions: 
\equ{
Q(\get_-) =  \frac 1{\sqrt{2k}}\,\Id,
\qquad Q'(\eta_-) = \frac{i \sqrt{k}}{\sqrt{2}} \, \Id,
\qquad R(\eta_-) = \Id.
\labl{initQ}
}
\item {\bf super-horizon ($\cH \mg k$)}: 
In this region we use $u$ to compute the vacuum correlator of 
the Newtonian potential $\gF$, which is related to $u$ via a simple 
rescaling, see \eqref{newvars}.  
As the $k^2$ dependence can be neglected, 
the exact solution for $u$ of equation \eqref{equ} is 
\equ{
\fl
u_{\vc k}(\eta) = u_{P\, \vc k}^{\;} + C_{\vc k} \gth
+ D_{\vc k} \gth 
\int_{\eta_\cH}^\eta \frac{\d\eta'}{\gth^2(\eta')},
\quad 
u_{P\, \vc{k}} = \gth \int^\eta_{\eta_\cH} \frac{\d\get'}{\gth^2}  
\int^{\eta'}_{\eta_\cH} \d \get'' \, \cH \gth \tget^\perp q_{2\, \vc{k}}, 
\labl{solu}
}
with $C_{\vc k}$ and $D_{\vc k}$ integration constants and $u_{P\, \vc k}$
a particular solution.
To work out $u_P$ in a more explicit form and to find solutions for $Q$ 
slow-roll assumptions are necessary, which are treated in section
\ref{slowrollpert}. 
}
As the sub-horizon region is irrelevant, what remains is the connection 
between the transition and the super-horizon region. 
In both these regions we have constructed analytic 
solutions of the same differential equation for $Q$. The only thing that must
still be computed to determine the super-horizon solution uniquely, is the 
relative overall normalization between the solutions in these two regions.
Instead of the more standard continuously differentiable matching at a
specific time scale, we do this by identifying leading order asymptotic 
expansions. 

This procedure works as follows. 
We can write both these solutions as power series in $k\eta$
and compare them in the transition region.
There we find that the leading powers of the transition and super-horizon 
solutions are the same, separately for both the 
decaying and the non-decaying independent solution. The ratio of the 
coefficients in front of these leading powers gives us the relative
normalization of the super-horizon solution with respect to the transition (and
sub-horizon) solution. Although we need to compute the coefficients accurately
to first order in slow roll, zeroth order turns out to be sufficient to 
distinguish the two independent solutions and identify the exponents of the 
leading terms in the expansions, see below \eqref{defEH}.  
To conclude, we can determine the solution valid in the super-horizon region 
uniquely from the solution in the transition region around $\eta_\cH$, even 
though the solution in the region in between is only known asymptotically. 
Some remarks on other matching schemes can be found at the end of 
section~\ref{pertsolcalc}.

\subsection{Slow roll for the perturbations}
\labl{slowrollpert}

To determine the solution for $Q$ in the super-horizon region, and to rewrite
the particular solution $u_P$ in terms of background quantities only, the
concept of slow roll on the perturbations is useful. We now justify the use of
slow roll on the perturbations and make this notion more precise.
Physically it represents the fact that the combination of background and 
perturbation modes far outside the horizon cannot be distinguished from the 
background.
We introduce the substitutions 
\equ{
\fl
\Bgf \ra \tilde{\Bgf} = \Bgf + \Bgd\Bgf, 
\qquad\qquad
b \ra \tb = a( 1 + \gF),
\qquad\qquad
a \ra \ta = a( 1 - \gF),
\labl{lin_proc}
} 
where we have chosen to work with conformal time after substitution
to make a direct comparison with section \ref{eqmotpert} 
possible. Notice that in this way the perturbed metric \eqref{separation} 
is obtained. Applying these substitutions to \eqref{eqmotback} and linearizing 
gives the perturbation equation \eqref{eqmotpert2} with $k^2$ put to zero, 
including the field curvature term. 
At the same time, by linearizing the combination
\equ{
\cD H_a + 3 H_a^2 - \gk^2 b^2 V = 0
\labl{Friedmanncomb}
}
of the Friedmann equations \eqref{Friedmann}, the equation of motion 
\eqref{eqPhi2} for $\gF$ is obtained.
In other words, for the super-horizon modes the 
system of background equations \eqref{eqmotbackiter} and \eqref{Friedmanncomb} 
for $(\Bgf,a,b)$ is also valid for the perturbed fields $(\tilde{\Bgf},\ta,\tb)$.
Hence the solutions for $(\Bgf,a,b)$ and $(\tilde{\Bgf},\ta,\tb)$ can only 
differ in their initial conditions, so that the perturbation quantities 
$(\Bgd\Bgf,\Phi)$ are obtained by linearizing the background quantities with 
respect to the initial conditions: 
\equ{
\Bgd\Bgf = (\Bnabla_{\Bgf_0} \Bgf) \Bgd\Bgf_0, 
\qquad\qquad
\mx{P}^\perp \vc{q} = a \, \mx{P}^\perp \Bgd\Bgf.
\labl{lininitcond}
}
This technique was also used in \cite{TaruyaNambu,SasakiTanaka}.
Here we have set the variations of the initial conditions $a_0$ and $b_0$ equal
to zero, as a simple counting argument shows that this is sufficient to generate
a complete set of solutions. 
Now if slow roll is valid for the background, it follows immediately that slow 
roll also governs the super-horizon perturbations. This fact has been used 
previously in the literature, see e.g.\ \cite{PolarskiStar,MukhanovSteinhardt}.

Applying slow roll to the equation of motion \eqref{eqtq} and rewriting it in
terms of the quantity 
\(
Q_{SR} \equiv \frac{a_\cH \sqrt{\tge_\cH}}{a \sqrt{\tge}} \, Q Q_\cH\inv,
\)
we find
\equ{
Q_{SR}' + \cH \left[ -\gd + (2\tge + \tget^\parallel)\Id + Z \right] Q_{SR} = 0,
\qquad\qquad
Q_{SR}(\eta_\cH) = \Id.
\labl{eqqSR}
}
Here we have used that $\cD \tQ = \tQ' - \cH \tQ$ and 
$\cD^2 \tQ = \tQ'' - 3 \cH \tQ' - (\cH' - 2\cH^2) \tQ$, 
because $\tQ$ scales with one power of $a$. 
For reasons that will become clear in the next section we have defined $\gd$ as
\equ{
\gd = - \frac{1}{3} \lh 2 \Id + \frac {\gO}{(1 - \tge)^2} \rh
= \tge \, \Id - \frac {\tM^2}{3H^2} + 2 \tge \, e_1^{\;} e_1^T,
\labl{defdelta}
}
where the second expression is valid to first order.
We make the additional assumption that also those components of
$\tM^2/H^2$ that cannot be expressed in terms of the slow-roll functions
defined in \eqref{slowrollfun} are of first order, so that $\gd$ is a 
first-order quantity.
The solution of \eqref{eqqSR} is found by integrating:
\equ{
Q_{SR}(\eta) = \exp \left [ \int^\eta_{\eta_\cH} 
\d\eta' \, \cH \lh \gd - (2\tge + \tget^\parallel)\Id - Z \rh \right ].
\labl{solQSR}
}
Although the initial conditions are applied at $\eta_\cH$, this solution is only
valid in the super-horizon region because $k^2$ terms have been neglected.
Since slow roll has been used, this result is a priori not expected to be
very accurate at the end of inflation. However, using the fact that the matrix 
between the brackets in \eqref{eqqSR} has its ($m1$) components ($m \geq 1$) all 
equal to zero to first order in slow roll (see \eqref{Me1} and \eqref{Zcomp}), 
one can easily find the solutions $(Q_{SR})_{11} = 1$ and $(Q_{SR})_{n1} = 0$ 
($n>1$). 
Since these vary slowly (or rather not at all) even at the end of inflation, 
slow roll is still a good approximation there for these components.

\subsection{Solutions of the perturbation equations to first order: calculation}
\labl{pertsolcalc}

In this section we perform the calculation that was discussed in
section~\ref{pertsolsetup}. As mentioned there,
in a sufficiently small interval around $\eta_\cH$ in the transition region
the slow-roll functions can be taken constant.
With this approximation we can obtain an expression for $\cH(\get)$ by 
integrating the relation for $\cH'$  in \eqref{slowrollrel}
with respect to conformal time, while integrating $N' = \cH$ 
gives the number of e-folds  to first order around $\eta=\eta_\cH$:
\equ{
\cH(\eta) = \frac{-1}{(1-\tge_\cH) \eta}, \qquad\qquad
N(\eta) = N_\cH - \frac 1{1 -\tge_\cH} \ln \frac{\get}{\get_\cH}.
\labl{apprx_hc}
}
Here we used the freedom in the definition of conformal time to set 
$\eta_\cH = -1/[(1-\tge_\cH) k]$. From \eqref{gthdblgth} we infer 
that to first order around $\eta=\eta_\cH$
\equ{
\gth(z) = \gth_\cH \lh \frac{z}{z_\cH} \rh^{1 +2 \tge_\cH +\tget^\parallel_\cH}, 
\qquad\qquad
z \equiv k \get,
\qquad 
\gth_\cH = \frac \gk{\sqrt{2}} \frac {H_\cH}{k \sqrt{\tge_\cH}}. 
\labl{gthtrans}
}
In these expressions we have made the conventional choice of $\eta_\cH$ as 
reference time to compute the constant slow-roll functions, etc., although in
principle one could do the complete computation with another reference time 
scale. However, to be able to take $\tgO$ constant (see below), this time
should not be much later than $\eta_\cH$.

With the initial condition \eqref{initQ} the solution of \eqref{eqtq} for the
rotation matrix $R$ during the transition region is
\equ{
R(z)= e^{ (N - N_-) Z_\cH} = 
\Bigl( \frac {z}{z_-} \Bigr)^{- \frac 1{1 - \tge_\cH} Z_\cH}. 
}
The only time dependent terms in the matrix $\gO$ \eqref{masseff} are first
order, so that we can take $\gO = \gO_\cH$ in the transition region.
The matrix $\tgO$ \eqref{eqtq} on the other hand is given by
\equ{
\fl
\tgO = R\inv(z) \gO_\cH R(z)
= \gO_\cH - [\gO_\cH,Z_\cH] \ln \frac{z}{z_-}
= \gO_\cH + 3 [\gd_\cH,Z_\cH] \lh \ln \frac{z}{z_\cH} 
+ \frac{3}{4} \ln \tge_\cH \rh,
\labl{rotgO}
}
where we used the definition of $\eta_-$ from the previous section and 
$\gd_\cH = \gd(\eta_\cH)$ is defined in \eqref{defdelta}.
In this section we are still considering a single, arbitrary mode $k$ 
(see equation \eqref{defetaH} and the text above equation \eqref{u1der}).
However, in the end we are interested in those modes that are visible in the
CMBR, which crossed the Hubble scale in a small interval about 60 e-folds before
the end of inflation. For those modes we estimate $\tge_\cH \sim 0.01$ 
(motivated for example by a quadratic potential, see
\eqref{approxtge}), so that $\ln \tge_\cH \sim \tge_\cH^{-1/2}$. Since both
$\gd$ and $Z$ are of first order, the time dependence of $\tgO$ caused by the
rotation is then only important at order 3/2 in the region around $z_\cH$. 
(For a smaller value of $\tge_\cH$ the effect
is even of higher order.) Hence we take $\tgO = \gO_\cH$. From the
correction term in equation \eqref{rotgO} we can always check explicitly if 
that assumption is justified.

For matching in the region around $z_\cH$ it will be useful to define 
\(
\bQ(z) \equiv R(z_\cH) \tQ(z).
\)
Then $Q(z) = \bQ(z)$ to first order in a sufficiently small region 
around $z_\cH$.
Using the same argument as in \eqref{rotgO} the corresponding 
$\bgO \equiv R(z_\cH) \tgO R\inv(z_\cH)$ is equal to $\gO_\cH$ to first 
order.
Using this result and equation \eqref{apprx_hc} for $\cH$, equation \eqref{eqtq} 
for $\tQ$ can be rewritten as an equation for $\bQ$:
\equ{
\bQ_{,zz} + \bQ - \frac {\gn_\cH^2 - \frac 14} {z^2} \, \bQ = 0, 
\qquad\qquad 
\gn_\cH^2 = \frac{9}{4} \Id + 3 \gd_\cH.
\labl{eqBessel}
}
The solution of this matrix equation can be written in terms of a Hankel 
function:\footnote{This Bessel equation and its solution in terms of Hankel
functions are well-known in the theory of inflationary density perturbations,
see e.g.\ \cite{boekLinde,MartinSchwarz} and references therein. However, in the
multiple field case under consideration the order $\nu$ of the Hankel function
is matrix valued. This should be considered in the usual way: defined by means
of a series expansion.}
\equ{
\bQ(z) =  \sqrt {\frac {\gp}{4k} }\, \sqrt {z} H_{\gn_\cH}^{(1)}(z),
\qquad\qquad\;\;
\gn_\cH = \frac{3}{2} \Id + \gd_\cH.
\labl{QsolHankel}
}
Here the initial conditions \eqref{initQ} at the beginning of the transition 
region have been taken into account, as can be seen by using the fact that for 
$|z| \mg 1$ the Hankel function can be approximated by 
$H_\gn^{(1)}(z) = \sqrt{ 2/ (\gp z)} \exp i( z - \gp \gn/2 -\gp/4)$ and
neglecting unitary matrices. 
We also need the leading order term in the expansion in $z$ of this result 
for $\bQ$:
\equ{
\bQ_{lo} = \frac{1}{i \sqrt{2\pi k}} \, \gG(\gn_\cH) 
\lh \frac{z}{2} \rh^{\half\Id-\gn_\cH}
= - \frac{e^{-i\pi\gd_\cH}}{i \sqrt{2 k}} \, E_\cH \lh \frac{z}{z_\cH} 
\rh^{-\Id-\gd_\cH}
\labl{Qtrz}
}
with 
\equ{
E_\cH \equiv (1- \tge_\cH) \Id  + (2 - \gg - \ln 2) \gd_\cH, 
\labl{defEH}
}
where $\gg \approx 0.5772$ is the Euler constant.
For later convenience we have defined the matrix $E_\cH$, which to zeroth order 
in slow roll is equal to the identity. In \eqref{Qtrz} we have taken only the
growing solution; the decaying one starts off with a term proportional to
$z^{\half\Id+\gn_\cH} = z^{+2\Id+\gd_\cH}$. We see that these two solutions 
can be distinguished already at zeroth order.

Next we turn to the super-horizon region. Here we have to relate $u$ and $Q$ 
by means of the first equation of \eqref{u1der}. The solution for $u$ is given 
in \eqref{solu}, while for $q_1$ we derive the following slow-roll equation of 
motion (see section~\ref{slowrollpert}):
\equ{
q_1' - \frac{(1/\gth)'}{1/\gth} \, q_1 = 2 \cH \tget^\perp q_2
\quad\Rightarrow\quad
q_1 = d \, \frac{1}{\gth} 
+ 2 \frac{1}{\gth} \int^\eta_{\eta_\cH} \d\eta' \, \cH \gth \tget^\perp q_2,
}
where we also gave the solution. By using slow roll we have selected the
non-decaying solution for $q_1/a$.
Using \eqref{u1der} we then find that the integration 
constant $D_\vc{k}$ in the solution \eqref{solu} for $u$ is given by
\(
D_\vc{k} = \half d.
\)
The constant $C_\vc{k}$ is irrelevant because the function $\gth$ rapidly
decays.
The integration constant $d$ can be determined using the procedure of
identification of leading order terms (leading order in the expansion in $z$,
not slow roll) described in section~\ref{pertsolsetup}.
Extrapolating the super-horizon solution for $q_1$ into the transition region
sufficiently close to $\eta_\cH$ that the integral can be neglected and that
$\gd_\cH \ln(z/z_\cH)$ is smaller than first order, and using \eqref{gthtrans} 
we find to first order
\(
q_1 = (d/\gth_\cH) (z/z_\cH)\inv.
\)
Under these conditions $e_1^T E_\cH (z/z_\cH)^{-\Id-\gd_\cH} 
= e_1^T E_\cH (z/z_\cH)\inv$ so that we can determine the constant $d$ from
equation \eqref{Qtrz}. (Notice that, as mentioned in section~\ref{pertsolsetup},
the exponents of $z/z_\cH$ need only be identified to zeroth order, so that
strictly speaking the condition that $\gd_\cH \ln(z/z_\cH)$ is smaller than
first order is not even necessary.)
The final first-order result for $D_\vc{k}$ is:
\equ{
\hat{D}_\vc{k} = - \half \frac{e^{-i\pi\gd_\cH}}{i \sqrt{2k}} \, \gth_\cH
e_1^T E_\cH \hat{a}^\dag + \mbox{c.c.}.
}
For later use we note that this identification procedure can also be used for 
the complete matrix $Q$, not just for $q_1$. Completely analogously to 
\eqref{solQSR} one has
\(
\bQ \propto a\, \exp \lh \int^\eta_{\eta_\cH} \d\eta' \, 
\cH \, R(\eta_\cH) R\inv \left [ \gd - \tge\Id \right ] R R\inv(\eta_\cH) \rh,
\)
which behaves like
\(
z^{-\Id-\gd_\cH}
\)
when extrapolated into the transition region around $\eta_\cH$ using 
\eqref{apprx_hc} and \eqref{rotgO}. Comparing with \eqref{Qtrz}
we see that it is exactly the leading order term in the expansion of the
Hankel function in the solution for $Q$ in the transition region that goes over 
into the dominant solution for $Q$ in the super-horizon region.

With this and \eqref{newvars} we can give the Newtonian potential $\gF$ as a 
quantum operator at late times during inflation up to first order in slow roll:
\equ{
\hat \gF_{\vc k}(t) = - \frac {\gk}{2 i k^{3/2}} \, 
\frac {H_\cH}{\sqrt{\tge_\cH}} \, 
\left(
A(t_\cH,t) \, e_1^T + \tU_P^T(t) 
\right) E_\cH e^{-i\pi\gd_\cH} \hat a^\dag_\vc{k} + \mbox{c.c.}.
\labl{opergF}
}
Here we used the identity $\cH_\cH = k$ and the functions $A(t_\cH,t)$ and 
$\tU_P^T(t)$ are defined as 
\equ{
A(t_\cH,t) = 1 - \frac{H(t)}{a(t)} \int_{t_\cH}^t \d t' a(t'),
\qquad
\tU_P^T = \frac{H}{a} \int^\eta_{\eta_\cH} \d\eta' a^2 \tge \, U_P^T(\eta'),
\non\\
U_P^T = 2 \sqrt{\tge_\cH} \int_{\eta_\cH}^{\eta} \d\eta' \, \cH
\frac{\tget^\perp}{\sqrt{\tge}} \, \frac{a_\cH}{a} \, e_2^T Q Q_\cH\inv.
\labl{defUPe}
}
In $A(t_\cH,t)$ we neglected one term which is exponentially suppressed with
the number of e-folds. In this and all following equations $Q_\cH$ is defined as
the leading order asymptotic expression for $Q$ evaluated at $\eta_\cH$, i.e.\
$Q_\cH = - E_\cH \exp(-i\pi\gd_\cH)/(i\sqrt{2k})$. Remember that $\gF_{\vc{k}}$
depends on $k$ not only explicitly, but also implicitly through the dependence
on $\eta_\cH$.

Using slow roll on the perturbations and substituting the result for $Q_{SR}$ 
from \eqref{solQSR} into the definition for $U_{P}^T$ we find
\equ{
U_{P}^T = 2 \int_{\eta_\cH}^{\eta} \d\eta' \, \cH
\tget^\perp \, e_2^T \exp \left [ \int^{\eta'}_{\eta_\cH} 
\d\eta'' \, \cH \lh \gd -(2\tge+\tget^\parallel)\Id - Z \rh \right ]
\labl{UPeinbg}
}
to first order in slow roll. This expression is given in terms of background
quantities only. $U_{P}$ has no component in the $e_1$ direction, since to first 
order $(Q_{SR})_{21} = 0$.
In section~\ref{flatmasses} we show how $U_{P}^T$ can be computed explicitly
for the case of a quadratic potential on a flat field manifold using the concept
of slow roll on the perturbations.

We have been able to determine the integration constant 
$D_{\vc k}$ in the solution for $u$ in the super-horizon region to first order
in slow roll by using analytic properties of the solutions for $Q$ in 
the transition region. We did not have to resort to a continuously 
differentiable matching at a specific time scale; the only time scale that
appears in the result is the reference time $\eta_\cH$ in the neighbourhood of
which we have expanded the solutions.
In the literature the concept of matching at a specific time is often used 
(see e.g.\ \cite{Mukhanovetal,MartinSchwarz}), for 
which usually the time of horizon crossing of either a generic or 
specific mode $k$ is used. 
On the one hand matching for the scales of observational interest is then 
performed at times when $k\eta \approx 1$, while on the other hand 
approximations only valid for small $k\eta$ are used. 
The identification procedure described in section~\ref{pertsolsetup} and used in
this section shows why the standard (single-field) results in the literature 
are nonetheless correct: by neglecting the $k$ dependent
corrections and taking $k\eta = 1$ (i.e.\ $z/z_\cH = 1$) one is exactly 
computing the overall normalization factor that we showed to be the only thing 
that needs to be determined. 
(Another possible way to solve the problem would be to match the
transition and super-horizon solutions at a specific time $\eta_+$ later than 
$\eta_\cH$, so that $k\eta_+ \ll 1$ is a valid assumption to first order.
However, it turns out that the interval between $\eta_\cH$ and this $\eta_+$ 
is too large to satisfy the requirement that the slow-roll functions can be 
taken constant, so that the Hankel solutions are not valid over the whole
interval to first order.)

\subsection{Adiabatic and isocurvature perturbations after inflation}
\labl{afterinfl}

The main subject of this paper is the treatment of perturbations during 
inflation, culminating in the result \eqref{opergF} for $\Phi$ for 
super-horizon modes during inflation. (The result for the other perturbations
$Q$ is given by \eqref{solQSR} with 
$Q_\cH = - E_\cH \exp(-i\pi\gd_\cH)/(i\sqrt{2k})$.)
However, in this section we discuss the relation between these results 
and the relevant quantities at the time of recombination when the CMBR was 
formed. Regarding the isocurvature perturbations after inflation we
follow the treatment by Langlois \cite{Langlois}, making several
simplifying assumptions. Other work on this subject can be found in e.g.\
\cite{KodamaSasakiiso,PolarskiStar,Bucheretal,Gordonetal,Bartoloetal} and
references therein.

Conventionally (see e.g.\ \cite{Gordonetal})  perturbations are divided into
adiabatic (or curvature) and entropy (or isocurvature) perturbations. Adiabatic
perturbations are perturbations in the total energy density, and are the only
ones present in the case of single-field inflation. If there are $N$ scalar
fields, there are in general $N-1$ isocurvature perturbations in addition to 
the single adiabatic perturbation. Isocurvature perturbations are perturbations 
in the ratios of energy densities of the different components, leaving the 
total energy density unperturbed. 
With our basis \eqref{basisconstr} the adiabatic perturbation
corresponds with the $e_1$ component, while the isocurvature perturbations
correspond in principle with all the other components. These different types of
perturbations can be sources for each other, see e.g.\ equation \eqref{equ}. By
including the particular solution in the result for $\Phi$ \eqref{opergF} we
have taken into account the effect of the isocurvature perturbations on the
adiabatic perturbation during inflation.

Although the treatment of purely adiabatic perturbations after inflation is 
rather straightforward, this is not the case for isocurvature perturbations. The
evolution of the isocurvature perturbations depends on some additional physical
input, most importantly to what kind of particles the multiple scalar fields
decay. If the interactions between these different kinds of particles are too 
strong, the isocurvature perturbations might not survive till the time of 
recombination at all \cite{Langlois}. We assume that one of the
fields decays to all the standard model particles, while the other fields decay
to different kinds of cold dark matter, so that there is no interaction between
the decay products of the different fields (except gravitational) and they can
be considered as ideal fluid components without mutual interactions. We also
neglect the effects of (p)reheating and assume that at the end of inflation
there is an immediate transition to a radiation dominated universe. For purely
adiabatic perturbations the presence of preheating is irrelevant on 
super-horizon scales, as was proved in \cite{Wandsetal}. In the presence 
of isocurvature perturbations preheating may have an important effect on the 
perturbations (see e.g.\ \cite{FinelliBrandenberger,HenriquesMoorhouse} and
references therein), but that is beyond the scope of this paper.

We first give the three vacuum correlators of the gravitational potential valid 
up to and including first order in slow roll at the time of recombination, and 
discuss their derivations afterwards. These three are the adiabatic
contribution (but including isocurvature effects during inflation), the
isocurvature contribution, and the mixing between them:
\equ{
\fl
\eqalign{
\langle \Phi_{\vc{k}\, ad}^2 \rangle_{t_{rec}} \;\;\;\: = \:  
\frac{9}{25} \, \frac{\gk^2}{4 k^3}
\frac{H_\cH^2}{\tge_\cH} \Bigl [ & (1-2\tge_\cH)(1 + U_{P\, e}^T U_{P\, e})\\
& + 2 B \lh (2\tge_\cH + \tget^\parallel_\cH)
+ 2 \tget^\perp_\cH e_2^T U_{P\, e} + U_{P\, e}^T \gd_\cH U_{P\, e} \rh \Bigr ],
\labl{CorrelatorPhi}
}}
\equ{
\fl
\langle \Phi_{\vc{k}\, iso}^2 \rangle_{t_{rec}} \;\;\;
= \frac{1}{25} \, \frac{\gk^2}{4 k^3} \frac{H_\cH^2}{\tge_\cH} 
\Bigl [ (1-2\tge_\cH) V_{e}^T V_{e}
+ 2 B V_{e}^T \gd_\cH V_{e} \Bigr ],
}
\equ{
\fl
\langle \Phi_{\vc{k}\, iso} \Phi_{\vc{k}\, ad} \rangle 
= \frac{3}{25} \, \frac{\gk^2}{4 k^3} \frac{H_\cH^2}{\tge_\cH} 
\Bigl [ (1-2\tge_\cH) V_{e}^T U_{P\, e}
+ 2 B \lh \tget^\perp_\cH e_2^T V_{e} + V_{e}^T \gd_\cH U_{P\, e} \rh \Bigr ],
}
with $U_{P\, e}^T \equiv U_P^T(\eta_e)$ given in \eqref{defUPe} and $V_e$ 
defined by
\equ{
V_e^T = \frac{1}{2} \sqrt{\tge_\cH} \, \frac{\sqrt{\tge_e} \, \tget^\perp_e}
{\tge_e + \tget^\parallel_e} \frac{a_\cH}{a_e} \, e_2^T Q_e Q_\cH\inv.
\labl{defVe}
}
Here $B \equiv 2 - \gg - \ln 2 \approx 0.7297$ and we used the definition of 
$E_\cH$ in \eqref{defEH} and the fact that $U_{P\, e}$ and $V_e$ have no 
components in the $e_1$ direction since $(a_\cH/a)(Q Q_\cH\inv)_{21}=0$, see 
below \eqref{solQSR}. 
Moreover, we assume $V_e$ to be of order 1 at most, otherwise 
other terms have to be included in the vacuum correlators to give the complete 
results to first order in slow roll. 

The explicit multiple field terms are the contributions of the terms 
$U_{P\, e}$ and $V_e$, which are absent in the single-field case. Since they 
are both to a large extent determined by $\tget^\perp$, we see that the 
behaviour of $\tget^\perp$ during the last 60 e-folds of inflation is crucial 
to determine whether multiple-field effects are important. 
For example, one can immediately draw the conclusion 
that in assisted inflation \cite{assinfl}, where one quickly goes to an 
attractor solution with all $\dot{\gf_i}$ equal to each other apart from 
constant factors, so that $\tget^\perp=0$, there will be no explicit 
multiple-field contributions to the gravitational potential.
One can draw the conclusion that with our basis the total isocurvature
perturbation that appears in the expressions for $\Phi$ only depends on 
the $e_2$ component of $q$, independently of the total number of 
fields and the actual number of independent entropy perturbations.
The fact that the entropy perturbations act as sources for the adiabatic
perturbation (the $U_{P}$ term) naturally leads to correlations between 
adiabatic and isocurvature perturbations, as described by the mixed correlator.
In fact these correlations are only absent if the source term disappears, 
i.e.\ again if $\tget^\perp$ vanishes. (The authors of \cite{Gordonetal} studied
the two-field case and found the derivative of the angle that
parameterizes the influence of the second field on the background trajectory to
be the relevant parameter. In the two-field limit this parameter corresponds
with $\tget^\perp$, but our result is valid for an arbitrary number of fields.)
In \cite{Langlois} these correlations were studied in the 
context of a double inflation model.

Using the concept of slow roll on the perturbations, introduced in
section~\ref{slowrollpert}, the quantity $U_{P\, e}$ can be rewritten in terms 
of background quantities only, as was done in \eqref{UPeinbg}.
Because we have used slow roll in the derivation, this expression is
in principle not valid at the very end of inflation. If \eqref{UPeinbg} does
indeed give a bad approximation for $U_{P\, e}$, for example if $\tget^\perp$
grows very large, a more careful treatment of the transition at the end of
inflation is necessary. However, in other cases the contribution to the integral
near the end of inflation can be negligible, for example if $\tget^\perp$ goes
sufficiently rapidly to zero at the end of inflation. In those cases
\eqref{UPeinbg} gives a very good approximation for $U_{P\, e}$ and the details
of the transition at the end of inflation are unimportant for the gravitational
correlator \eqref{CorrelatorPhi}. An important example of this latter case
is discussed in section~\ref{flatmasses}.
Unfortunately $V_e$ depends very much on the details of the transition at the 
end of inflation, so that for an actual calculation a model of this transition 
has to be assumed, a treatment of which is beyond the scope of this paper.

We conclude this section by discussing the derivation of the correlators.
Assuming the absence of anisotropic stress, the equation for super-horizon
modes of $\Phi$ derived from the $(00)$ and $(ij)$ components of the Einstein
equations  can be written as (see e.g.\ \cite{Mukhanovetal,HwangNoh}):
\equ{
\Phi'' + 3 \cH(1+c_s^2) \Phi' + \gk^2 a^2 (\gr c_s^2 - p) \Phi
= \half \gk^2 a^2 e
\labl{Phiafter}
}
with 
\equ{
\fl
\gr \equiv - T^0_{\;0} = \frac{3}{\gk^2 a^2} \, \cH^2, 
\quad
p \equiv \frac{1}{3} T^i_{\;i} = - \frac{1}{\gk^2 a^2} (2 \cH' + \cH^2),
\quad
c_s^2 \equiv \frac{p'}{\gr'}, 
\quad
e \equiv \gd p - c_s^2 \gd\gr.
}
Here $\gr$ and $p$ are the energy density and isotropic pressure, $c_s^2$ is the
sound velocity and $e$ is the total (gauge invariant) entropy perturbation.
The expressions after the second equality sign in the definitions of $\gr$ and
$p$ follow from the background Einstein equations. The perturbations are
analogously defined as $\gd\gr\equiv-\gd T^0_{\;0}$ and 
$\gd p\equiv\frac{1}{3}\gd T^i_{\;i}$.

The adiabatic perturbation at the time of recombination is usually defined
as the homogeneous solution of \eqref{Phiafter} at that time,
where as initial conditions one has matched to the total solution for $\Phi$ at
the beginning of the radiation dominated era. Hence the adiabatic perturbation
does include the effects of isocurvature perturbations during inflation (and in
general preheating). We can rewrite the homogeneous part of equation
\eqref{Phiafter} as $u''-(\gth''/\gth)u=0$ with $u$ and $\gth$ given by the same
definitions \eqref{newvars} and \eqref{gthdblgth}, which are also 
defined after inflation in terms of $a$ and its derivatives denoted by $\cH$ and
$\tge$. Hence we can use the same solution \eqref{solu}, with different 
constants $\tC_{\vc k}$ and $\tD_{\vc k}$ and without the particular solution. 
Matching our complete solution for $\Phi$ at the end of inflation $t_e$ to the 
homogeneous solution for $\Phi$ after inflation, we find for the latter:
\equ{
\eqalign{
\Phi_{\vc{k}\, ad}(t) \: & = \: 2 \lh D_\vc{k} + \gth_e \, 
u_{P \, \vc{k}}'(\eta_e) - \gth'_e \, u_{P \, \vc{k}}(\eta_e) 
\rh A(t_e,t)\\
& = - \frac{\gk}{2 i k^{3/2}} \frac{H_\cH}{\sqrt{\tge_\cH}} A(t_e,t)
\lh e_1^T + U_{P\, e}^T \rh E_\cH e^{-i\pi\gd_\cH} \hat{a}_{\vc{k}}^\dag 
+ \mbox{c.c.}
}
\labl{Phiadafter}
}
This expression is only valid some time after $t_e$, since we have neglected
the $\tC_{\vc k}$ term, which is suppressed by $H/a$. The function $A$ is 
defined in \eqref{defUPe}.
At the time of recombination we can use that $a(t) \propto t^{2/3}$ in a matter
dominated universe to find $A(t_e,t_{rec}) = 3/5$.\footnote{Sometimes the
curvature perturbation on super-horizon scales 
$-\gz=\Phi-(H/\dot{H})(\dot{\Phi}+H\Phi)$ is used in the literature instead 
of the gravitational potential (see e.g.\ \cite{Gordonetal} and references
therein), since it is constant for purely adiabatic perturbations
\cite{Wandsetal}. However, as one can see from \eqref{Phiadafter}, this
definition simply removes the time dependent factor $A$. Because of the
expansion of the universe $A$ has a well-defined value at recombination,
independent of evolutionary details, so that one can use $\Phi$ as well.}

The isocurvature contribution
to the gravitational potential at the time of recombination is usually defined
as the particular solution of \eqref{Phiafter}, with the initial conditions that
it is zero and has zero derivative at the beginning of the radiation dominated
era. Following \cite{Langlois} we assume that $N-1$ fields have decayed to 
non-interacting cold dark matter (ideal fluid) components and the remaining
field to the standard model particles, which we represent by the photons and 
denote by the subscript $r$. Then we can write (using \cite{HwangNoh}):
\equ{
\tS \equiv -\frac{1}{4} \frac{e}{\gr c_s^2 - p} 
= \frac{\sum_{i=1}^{N-1} \gr_i S_i}{\sum_{i=1}^{N-1} \gr_i},
\labl{deftS}
}
where $S_i$ is defined by 
\(
S_i \equiv \gd\gr_i/(\gr_i+p_i) - \gd\gr_r/(\gr_r+p_r)
\)
and we used $p_i=0$ for the CDM components and $p_r=\gr_r/3$ for the photons. 
In \cite{KodamaSasakiiso,HwangNoh} it is proved that for the case of ideal 
fluid components
without mutual interactions in a flat universe, the super-horizon modes of $S_i$
are constant. Since the $\gr_i$ all have the same time dependence 
$\propto a^{-3}$, this means that $\tS$ is also constant. Hence we find a very
simple particular solution for $\Phi$: $\Phi_p = -2 \tS$. To take care of the
initial conditions we have to add a part of the homogeneous solution and find
\equ{
\Phi_{iso}(t) = -2 \tS + 2 \tS \lh 1 + \frac{1}{\tge_e} \rh A(t_e,t).
\labl{Phiiso}
}
Since by assumption $\tge_e = \tge_{rad} = 2$ and $A(t_e,t_{rec})=3/5$ we find
at the time of recombination $\Phi_{iso}(t_{rec}) = -\tS/5$.
To link $\tS$ to inflationary quantities one computes $\tS$ according to its 
definition \eqref{deftS} during inflation and makes the assumption that
$\tS$(after inflation) $=\tS$(end of inflation). The result is:
\equ{
\tS = \frac{\gk}{2\sqrt{2}} \left [ \frac{\sqrt{\tge}}{\tge+\tget^\parallel} 
\, \tget^\perp \, \frac{q_2}{a} \right ]_e.
\labl{tSinfl}
}
This is not a slow-roll approximated expression; the slow-roll functions 
are merely short-hand notation defined by \eqref{slowrollfun}. 


\section{Illustration: scalar fields with a quadratic potential}
\labl{flatmasses}

\subsection{Analytical expressions for background and perturbations}
\labl{analytics}

In this section we consider slow-roll inflation with scalar 
fields living on the flat manifold $\Real^N$ with a quadratic potential $V$.
The slow-roll equation of motion and Friedmann equation for the background 
quantities to first order are given by
\equ{
\fl
\Dot \Bgf = - \frac 2{\sqrt{3}\, \gk} \, \Bder^T \sqrt{V(\Bgf)},
\qquad
H = \frac{\gk}{\sqrt{3}} \, \sqrt{V(\Bgf)} \lh 1 + \frac{\tge}{6} \rh, 
\qquad
V = \half \gk^{-2} \, \Bgf^T \Bsfm^2 \Bgf.
\labl{emSRflat}
}
Here $\Bsfm^2$ is a general symmetric mass matrix given in units 
of the Planck mass $\gk^{-1}$.  The initial starting point of the field $\Bgf$ 
is denoted by $\Bgf_0 = \Bgf(0)$.
The solution of the equation of motion \eqref{emSRflat} can be written in 
terms of one dimensionless positive scalar function $\gps(t)$:
\equ{
\fl
\Bgf(t) = e^{ - \half \Bsfm^2 \gps(t)} \Bgf_0
\qquad \Ra \qquad 
\dot{\gps} = \sqrt{\frac {2}{3}} \frac{2}{\gk^2 \gf_0}
\left(  
\hat{\Bgf}_0^T \Bsfm^{2} e^{-\Bsfm^2 \gps} \hat{\Bgf}_0
\right)^{-\half},
\labl{ansatz_Bgf}
}
with the initial condition  $\gps(0)=0$ and where
$\hat{\Bgf}_0 \equiv \Bgf_0/\gf_0$, with $\gf_0 \equiv |\Bgf_0|$, denotes the 
unit vector in the direction of the initial position in field space. 
In other words, we have determined the trajectory that the field $\Bgf$ 
follows through field space starting from point $\Bgf_0$. 
The number of e-folds $N=\int H \d t$ and the slow-roll function $\tge$ can 
be given as a function of $\gps$ by using \eqref{ansatz_Bgf}: 
\equ{
\eqalign{
N(\gps) = 
N_\infty \lh 1 - \hat{\Bgf}_0^T e^{-\Bsfm^2 \gps} \hat{\Bgf}_0 \rh
- \frac{1}{12} \ln \frac{\hat{\Bgf}_0^T \Bsfm^2 e^{-\Bsfm^2 \gps} 
\hat{\Bgf}_0}{\hat{\Bgf}_0^T \Bsfm^2 \hat{\Bgf}_0},\\
\tge(\gps) = \frac 1{2N_\infty} 
\frac {\hat{\Bgf}_0^T \Bsfm^4 e^{-\Bsfm^2 \gps} \hat{\Bgf}_0}
{( \hat{\Bgf}_0^T \Bsfm^2 e^{-\Bsfm^2 \gps} \hat{\Bgf}_0)^2}
\labl{Ntge}
}}
with $N_\infty =  \frac{1}{4} \gk^2 \gf_0^2.$
For the other slow-roll functions similar expressions can be obtained. 
The slow-roll limit for the total number of e-folds during inflation
$N_\infty$ is approached by taking the limit $\gps\ra\infty$ in the zeroth order
expression for $N$, i.e.\ the above expression without the logarithm. 

It is useful to have a leading order estimate of $\tge_\cH$. To this end we take
the zeroth order expression for $N(\psi)$ and assume that $\psi$ is already so
large at time $t_\cH$ that we can neglect all masses except the smallest one in
the exponential $\exp(-\half\Bsfm^2\psi)$. Then we can solve for $\psi_\cH$ and
insert this into the expression for $\tge(\psi)$ to find
\equ{
e^{-m_1^2 \gps_\cH} = \frac{1}{||\mx{E}_1||^2} \frac{N_\infty-N_\cH}{N_\infty}
\qquad\Rightarrow\qquad
\tge_\cH = \frac{1}{2 (N_\infty-N_\cH)}.
\labl{approxtge}
}
Here $m_1$ is the smallest mass eigenvalue, $\mx{E}_1$ is the projection 
operator that projects on the eigenspace of $m_1$ and 
$||\mx{E}_1||^2 = \hat{\Bgf}_0{}^T \mx E_1 \hat{\Bgf}_0 \leq 1$. 
Since $N_\infty-N_\cH \approx 60$ we see that $\tge_\cH \sim 0.01$.

We continue by computing the particular solution $U_{P\, e} = U_P(\eta_e)$ 
defined in \eqref{defUPe}. It turns out that in this case we can work out the 
integral analytically in slow roll, making use of the fact that we have 
obtained the slow-roll trajectories in \eqref{ansatz_Bgf}. 
The velocity and acceleration are given by
\equ{
\Dot \Bgf = - \half \dot \gps \, \Bsfm^2 \Bgf, 
\qquad\qquad
\cD_t \Dot \Bgf = - \half \ddot \gps \, \Bsfm^2 \Bgf + 
\frac 14  \dot \gps^2 \, \Bsfm^4 \Bgf, 
\labl{velo_acc}
}
while according to \eqref{lininitcond} we obtain $\Bgd\Bgf$ by varying $\Bgf$ 
with respect to the initial conditions:  
\equ{
\Bgd\Bgf = - \half \gd \gps \, \Bsfm^2 \Bgf 
+ e^{-\half \Bsfm^2 \gps} \Bgd \Bgf_0,
\labl{vary_Bgf}
}
where $\gd\gps$ is the function $\gps$ varied with respect to $\Bgf_0$. 
The projector parallel to the velocity is given by
\(
\mx P^\parallel = {\Bsfm^2 \Bgf\, \Bgf^T \Bsfm^2}/
({\Bgf^T \Bsfm^4 \Bgf}),
\)
and therefore we find that 
\equ{
\cD_t \Dot\Bgf^T \, \mx P^\perp \Bgd\Bgf = 
\frac 14 \dot\gps^2 \, \Bgf^T 
\left[
\Bsfm^4 
- \frac {\Bgf^T \Bsfm^6 \Bgf}{\Bgf^T \Bsfm^4 \Bgf} \Bsfm^2
\right] e^{-\half \Bsfm^2 \gps} \Bgd \Bgf_0.
}
Here we have used that the first terms of $\cD_t \Dot \Bgf$ 
and $\Bgd\Bgf$ are proportional to $\Dot \Bgf$ and hence are projected 
away, so that $\gd\gps$ drops out. 
We rewrite $U_{P\, e}^T$ such that we can apply this result: 
\equ{
U_{P\, e}^T q_\cH  
= 2 \sqrt {\tge_\cH} \,  \int_{t_\cH}^{t_e} \d t\, \frac{H}{\sqrt \tge} \, 
\tilde{\Bget}^T  \mx P^\perp \, a_\cH \Bgd\Bgf.
}
Substituting the definition \eqref{slowrollfun} for $\tilde{\Bget}$ and using 
\eqref{Ntge} for $\tge$ and \eqref{velo_acc} to determine $|\Dot\Bgf|$,
the integral takes the form 
\equ{
U_{P\, e}^T q_\cH = \frac{\gk\sqrt{\tge_\cH}}{\sqrt{2}} \int_{\gps_\cH}^{\gps_e} 
\d \gps \, \frac {\Bgf^T \Bsfm^2 \Bgf}{\Bgf^T \Bsfm^4 \Bgf}
\, \Bgf^T \Bsfm^4 \mx P^\perp 
e^{-\half \Bsfm^2 (\gps-\gps_\cH)} \, a_\cH \Bgd \Bgf_\cH.
}
By writing out the projector $\mx P^\perp$ we can employ 
\equ{
\frac 1{\Bgf^T \Bsfm^4 \Bgf}
\left[
\Bgf^T \Bsfm^4 
- \frac {\Bgf^T \Bsfm^6 \Bgf}{\Bgf^T \Bsfm^4 \Bgf} \Bgf^T \Bsfm^2
\right] e^{-\half \Bsfm^2 \gps}  = 
- \pp[]{\gps} 
\left[
\frac {\Bgf^T \Bsfm^2 e^{-\half \Bsfm^2 \gps}}
{\Bgf^T \Bsfm^4 \Bgf}
\right]
}
to perform a partial integration to express $U_{P\, e}^T$ as  
\equ{
U_{P\, e}^T q_\cH = \frac{\gk\sqrt{\tge_\cH}}{\sqrt{2}}  
\left[ \Bgf^T \mx P^\perp e^{-\half \Bsfm^2(\gps-\gps_\cH)}
\right]_{\gps_\cH}^{\gps_e} \,a_\cH \Bgd\Bgf_\cH.
\labl{UPE}
}
To determine $a_\cH \gd\gf_\cH$ we use the definition of $q$ in \eqref{newvars}:
\(
q_\cH = a_\cH ( \gd \gf_\cH + (\sqrt{2 \tge_\cH}/\gk) \, \gF_\cH e_1 ), 
\)
where we also inserted the definition of $\tge$.
Using \eqref{newvars} and \eqref{uindgv} to relate $\Phi_\cH$ to $q_\cH$
we obtain
\equ{
\fl
a_\cH \gd \gf_\cH = X^T q_\cH, \qquad\mbox{with}\qquad
X = 1 - \tge_\cH \lh 2 \tget^\perp_\cH e_2^{\,} 
+ \lh 2 \tge_\cH + \tget^\parallel_\cH - \gd_\cH \rh 
e_1^{\,} \rh e_1^T,
\labl{gdgf+}
}
where we made use of the relation $Q'_\cH = \cH_\cH(1-\tge_\cH+\gd_\cH)Q_\cH$
that follows from \eqref{Qtrz}. Hence $X=1$ to first order in slow roll.
With this we find our final result for $U_{P\, e}^T$:
\equ{
U_{P\, e}^T = \frac{\gk \sqrt{\tge_\cH}}{\sqrt{2}}
\Bigl( - \gf_\cH^\perp + e^{-\half {\mathsf{m}}^2(\gps_e-\gps_\cH)}
\gf_e^\perp \Bigr)^T.
\labl{UPevb}
}
Here all terms are written in terms of the basis $\{\vc{e}_n\}$: $\gf^\perp$
denotes the vector with components $\vc{e}_n^\dag \mx{P}^\perp \Bgf$ and 
$\exp(-{\mathsf{m}}^2 \gps)$ the matrix with components
$\vc{e}_m^\dag \exp(-\Bsfm^2 \gps) \vc{e}_n$.
The second term within the parentheses in the expression for $U_{P\, e}^T$ is in
general very small. In the first place all but the least massive field will 
have reached zero near the end of inflation, so that $\gf_e^\perp$
is small. In the second place this term is suppressed by the large negative
exponential, since $\gps_e$ is very large near the end of inflation, even though
we may not be able to take the limit of $\gps_e \rightarrow \infty$ since slow
roll is then not valid anymore. 

\subsection{Numerical example}
\labl{numerics}

\begin{figure}
\begin{center}
\epsfxsize=7.0cm \epsfbox{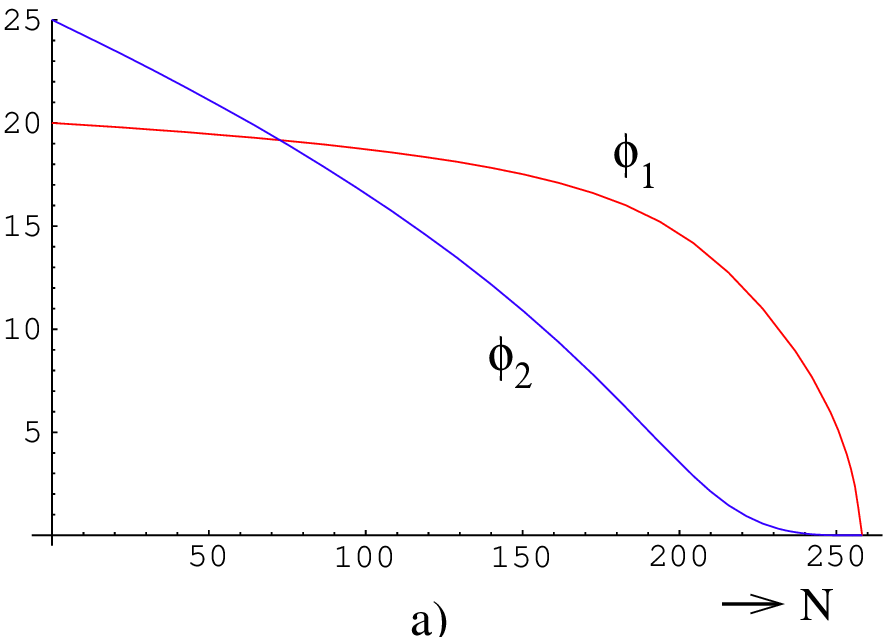}
\hspace{0.5cm}
\epsfxsize=7.5cm \epsfbox{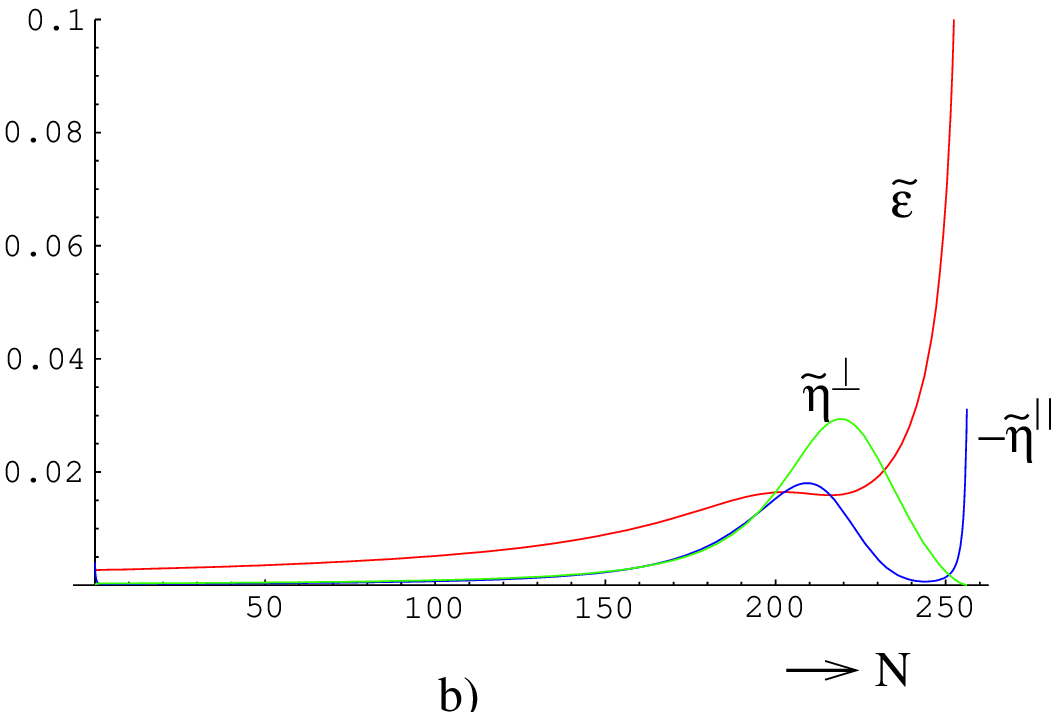}
\end{center}
\caption[fig1]{\labl{fieldfig} a) Background fields and b) slow-roll functions 
as a function of the number of e-folds in the model with two fields on 
a flat manifold with a quadratic potential with masses $m_1 = 1 \cdot 10^{-5}$, 
$m_2 = 2.5 \cdot 10^{-5}$ and initial conditions $\gf_1 = 20$, $\gf_2 = 25$.}
\end{figure}

We now treat a numerical example, not only to illustrate the theory, but also to
check our analytical results. We take the situation of two fields, with masses
$m_1 = 1 \cdot 10^{-5}$ and $m_2 = 2.5 \cdot 10^{-5}$ in units of the Planck
mass. As initial conditions we choose $\gf_1 = 20$ and $\gf_2 = 25$, also in
Planckian units. Then $N_\infty = 256.25$, while an exact numerical calculation
gives a total amount of inflation of $257.8$ e-folds before the oscillations 
start. We have chosen the overall normalization of the masses such that we get 
the correct order of magnitude for the amplitude of the density perturbations.
Apart from giving sufficient inflation, the specific choice of initial 
conditions has no special meaning.
We compute all background quantities exactly, as we want to check the accuracy 
of our analytical results for the perturbations. In figure~\ref{fieldfig}
we have plotted the fields and slow-roll functions as a function of
the number of e-folds. We see that the more massive field goes to zero more
quickly than the less massive field, as expected from \eqref{ansatz_Bgf}.
Moreover, around the time that the second field reaches zero, all slow-roll
functions show a bump. For the chosen masses and initial conditions the bumps 
are located during the last 60 e-folds.
As mentioned in the paragraph below \eqref{defVe}, for multiple-field effects 
to be important we need $\tget^\perp$ to be substantial during the last 60 
e-folds. Hence this is a good model to look for multiple-field effects.
Moreover, as we see from the figure, $\tget^\perp$ goes to zero at the end of
inflation, so that we expect corrections to $U_{P\, e}$, caused by the 
break-down of slow roll at the end of inflation, to be small. Indeed, 
figure~\ref{UPfig} shows that the contribution to $U_{P\, e}$ during the last 
few e-folds of inflation is negligible.

\begin{figure}
\begin{center}
\epsfxsize=7.0cm \epsfbox{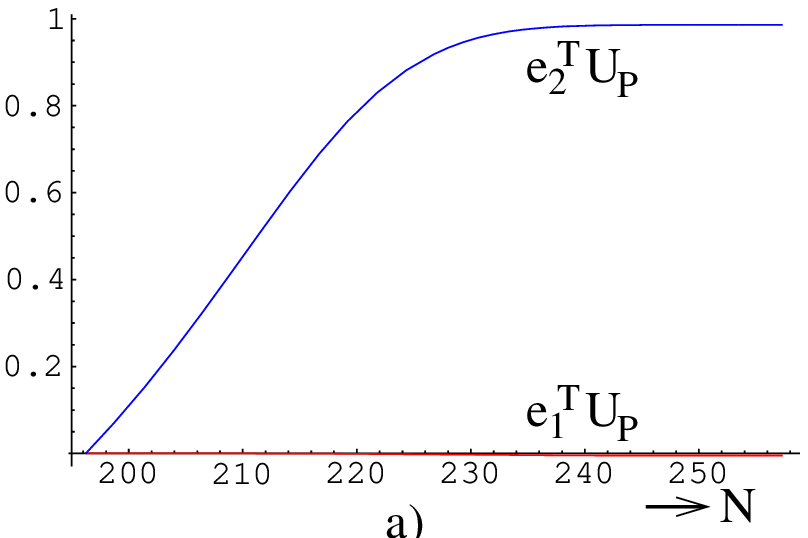}
\hspace{0.5cm}
\epsfxsize=7.0cm \epsfbox{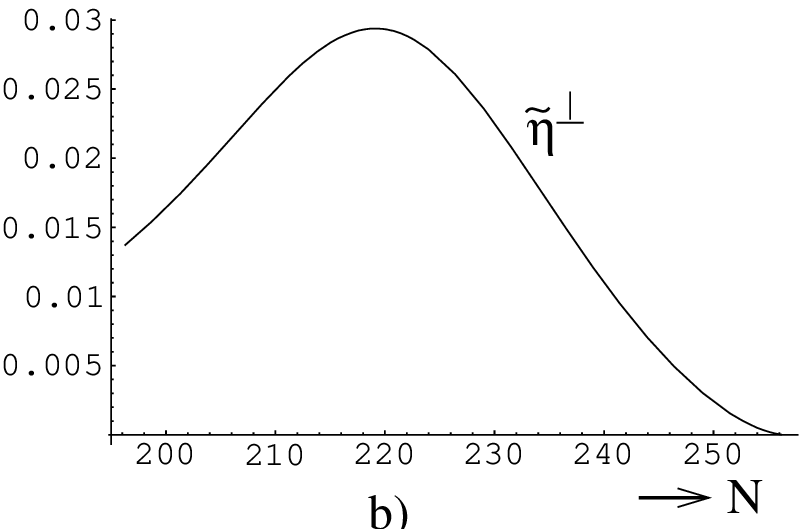}
\end{center}
\caption[fig2]{\labl{UPfig} a) The particular contribution $U_{P}$ to the 
gravitational correlator during the super-horizon region as a function of the 
number of e-folds. To show the
relation with the behaviour of $\tget^\perp$, this slow-roll function has been
plotted again in figure b), on the same horizontal scale as figure a).} 
\end{figure}

The results for the amplitude of the adiabatic vacuum correlator of the 
gravitational potential are summarized in table \ref{tabel}. We split the 
contributions to the correlator into a homogeneous part (all terms without 
$U_{P\, e}$) and a particular part (the rest, so including mixing terms). 
Everything is evaluated for the mode $k$ that crossed the horizon 60 e-folds
before the end of inflation.
The last column gives the relative error between our first order analytical
results \eqref{CorrelatorPhi} and \eqref{UPevb} on the one hand, and the exact
numerical result on the other. These results agree with our claim that we
computed the correlator to first order in slow roll: the relative errors are
(much) smaller than $\cO(\tge_\cH)$. We also see that our slow-roll 
approximation for $U_{P}$ is indeed still very good at the end of inflation.
The column before that shows the relative contributions of the various parts to
the total adiabatic correlator. We see that the particular solution terms are 
responsible for almost half the total result in this model. Hence neglecting 
these terms to leading order, which might naively be done because they couple 
with a $\tget^\perp$ in \eqref{equ}, can be dangerous.

\begin{table}
\caption[tabel]{\labl{tabel} The amplitude of the adiabatic vacuum 
correlator of the gravitational potential
\(
|\gd_\vc{k}|^2 = \frac{1}{2\pi^2} k^3 \langle \Phi_{\vc{k}\, ad}^2 (t_{rec})
\rangle 
\)
for the mode $k$ that crossed the horizon 60 e-folds before the end of
inflation, is separated into a purely homogeneous and a (mixed) particular part.
The first two columns give their values and their relative contributions to 
the total correlator according to our analytical slow-roll result 
\eqref{CorrelatorPhi} combined with \eqref{UPevb}. The final column shows the 
relative error between these expressions and the exact numerical results.} 
\begin{indented}
\item[] \tabu{@{}lccc}{
\br
& Amplitude $|\gd_\vc{k}|^2$ & Contribution to total & Relative error\\ 
\mr
Homogeneous & $1.55 \cdot 10^{-9}$ & 0.505 & 0.0001\\
Particular & $1.52 \cdot 10^{-9}$ & 0.495 & 0.0006\\ 
\mr
Total & $3.08 \cdot 10^{-9}$ & 1 & 0.0003\\
\br
}
\end{indented}
\end{table}

For (significantly) larger or smaller mass ratios, $\tget^\perp$ is smaller
during the last 60 e-folds and the contribution of the explicit
multiple-field terms to the correlator is less important. This could also
be expected a priori, since a much larger mass ratio means that the heavy field
has already reached zero before the last 60 e-folds, and the situation is
effectively single-field. On the other hand, a much smaller mass ratio means
that we approach the limit of equal masses, which corresponds with a central
potential that is also effectively single-field.

\section{Conclusions and discussion}
\labl{conclusions}

We have given a general treatment for scalar perturbations on a flat 
Robertson-Walker spacetime in the presence of an arbitrary number of scalar 
fields that take values on a curved field manifold during slow-roll inflation. 
These are the kind of systems that one typically obtains from (string-inspired)
high-energy models.
The scalar perturbations are calculated to  first order in slow roll. In 
particular we compute the vacuum correlator of
the gravitational potential in terms of background quantities  only, which is
related to the temperature fluctuations that are observed in the CMBR.

A discussion of the background scalar fields served  as the foundation for this
analysis.  The first of three central ingredients for this discussion  is the
manifestly covariant treatment with respect to reparameterizations  of the
field manifold and of the time variable.  Secondly, the field dynamics (the
field velocity, acceleration, etc.) naturally induce an orthonormal basis ($\vc
e_1, \vc e_2,\ldots$) on the field manifold.  This makes a separation between
effectively single-field and truly multiple-field  contributions possible.
Finally, we modified the definitions of the well-known  slow-roll parameters to
define slow-roll functions in terms of derivatives of  the Hubble parameter and
the background field velocity for the case of multiple scalar field inflation.
These slow-roll functions are vectors, which can be decomposed in  the basis
induced by the field dynamics. For example, the slow-roll function 
$\tget^\perp$ measures the size of the acceleration perpendicular to the field 
velocity. Because we did not make the assumption that slow roll is valid in the
definition of the slow-roll functions,  it is often possible to identify these
slow-roll functions in exact equations of motion and make decisions about
neglecting some of the terms. However, more important for precision
calculations are  estimates of the accuracy of the solutions of these
approximated slow-roll equations; it turns out that if the size of the region
of integration is too large this accuracy may be compromised. 

Our calculation of the scalar perturbations accurate to first order  in slow
roll is based on the following cornerstones.  We generalized the combined
system of gravitational and matter  perturbations of Mukhanov et al.\
\cite{Mukhanovetal} by defining   the Mukhanov-Sasaki variables 
as a vector  on the scalar field manifold. The
decomposition of these variables in the basis  induced by the background field
dynamics is  field space reparameterization invariant, and the corresponding
Lagrangean  takes the standard canonical form, making quantization
straightforward. The  gravitational potential only couples  to the scalar field
perturbation in the direction $\vc e_2$ with a slow-roll  factor
$\tget^\perp$. 

To obtain analytic solutions for the scalar perturbations to first order 
in slow-roll, it is crucial to divide the inflationary epoch into three 
different regimes, which reflects the change of behaviour for a given mode when
it crosses the Hubble scale. These regimes are conventionally
called sub-horizon, horizon-crossing (transition), and super-horizon. 
Within all three regions analytic solutions for the perturbations
valid to first order could be found. The sub-horizon region is irrelevant for 
the correlator of the gravitational potential. Relating the transition and 
super-horizon regions is not trivial, as there is no analytic result that is 
valid to first order at the boundary between them. Using the procedure 
where we identify leading order asymptotic expansions, we could 
determine the relative normalization of the super-horizon solution with respect
to the solution in the sub-horizon region using analytic properties.

To determine the solution for the scalar perturbations other than the
gravitational potential in the super-horizon region we need a final 
cornerstone: the application of slow roll  to the perturbations.
For this it was essential that we treated the background  using an arbitrary 
time variable, since the perturbed metric has to be rewritten in terms of a
changed background metric. In particular this method was used to obtain an 
integral expression for the particular solution of the gravitational potential 
in terms of background quantities only.
Although this expression is a priori not expected to be good near the end of
(slow-roll) inflation, we show that it can actually be a very good approximation
if $\tget^\perp$ goes to zero at the end of inflation.

Making some assumptions about the evolution of the universe and its
matter content after inflation we gave first-order expressions for the
adiabatic, isocurvature and mixing correlators of the gravitational potential
at the time of recombination. Here we neglected the possible effects of
(p)reheating, which are still under investigation. The adiabatic correlator
includes the effect of entropy perturbations acting as a source for the
gravitational potential during  inflation and can be given in terms of
background quantities expressed during the transition region (apart from
possible end-of-inflation effects in the particular solution which are absent
if $\tget^\perp$ goes to zero). The isocurvature correlator depends strongly on
the transition at the end of inflation.

Finally, we discussed the example of multiple scalar fields on a flat manifold
with a quadratic potential. To first order the trajectory of all fields 
through field space can be found in terms of one function of time, 
and the particular solution can be determined completely analytically using 
the slow-roll approximation on the perturbations. We concluded with 
an explicit numerical check of the amplitude of the adiabatic correlator and 
found this to be consistent with our analytical results.

Multiple-field effects are important in the adiabatic correlator of the 
gravitational potential if $\tget^\perp$ is sizable during the last 60 e-folds 
of inflation. The most important source of multiple-field effects is the 
particular solution of the gravitational potential. We found in our numerical 
example that this term can contribute even at leading order. Hence it can be 
dangerous to neglect this term, even when looking only at leading order. This 
contribution is included implicitly in the function $N(\Bgf)$ of 
\cite{NakamuraStewart}, but we derived an explicit expression. 
If $\tget^\perp$ peaks in the transition region, the rotation of the basis 
induced by the background field dynamics over the transition region can be 
another source of multiple-field effects, but in generic situations we found 
it to be beyond the level of first order in slow roll.
Although the details of the isocurvature and mixing correlators of the 
gravitational potential depend on the transition at the end of inflation, an
important role is again played by the slow-roll function $\tget^\perp$. If it is
zero there will be no mixing between adiabatic and isocurvature perturbations.
Moreover, we found that 
with the assumption of an immediate transition to a
radiation dominated universe at the  end of inflation, 
only the $e_2$ component
of $q$ (using our basis) enters  into the final expression for the isocurvature
correlator, independently of the  actual number of entropy perturbations.

\ack

This work is supported by the European Commission RTN
programme HPRN-CT-2000-00131.
S.G.N.\ is also supported by  
piority grant 1096 of the Deutsche Forschungsgemeinschaft
and European Commission RTN programme HPRN-CT-2000-00148/00152.

\appendix

\section{Geometrical concepts}
\labl{geometry}

Consider a real manifold $\cM$ with metric $\mx G$ and local coordinates 
$\Bgf = (\gf^a)$. From the components of this metric $G_{ab}$ 
the metric-connection $\gG^a_{bc}$ is obtained using the metric postulate. 
The curvature tensor of the manifold can be introduced 
using tangent vectors $\vc{B}, \vc{C}, \vc{D}$:
\equ{
\fl
[\mx{R}(\vc{B},\vc{C})\vc{D}]^a \equiv
R^a_{\; bcd} \, B^b C^c \, D^d \equiv
\lh \gG^a_{bd,c} - \gG^a_{bc,d} + \gG^e_{bd} \gG^a_{ce}
- \gG^e_{bc} \gG^a_{de} \rh  B^bC^c \, D^d.
}
One should realize that for notational convenience we do not use the 
standard definition as made for example in \cite{Nakahara}: 
our $\mx{R}(\vc{B},\vc{C})\vc{D}$ is conventionally 
denoted by $\vc{R}(\vc{C},\vc{D},\vc{B})$.

The metric $\mx{G}$ introduces an inner product and the 
corresponding norm on the tangent bundle of the manifold: 
\equ{
\vc{A} \cdot \vc{B} = 
\vc A^\dag \vc B \equiv 
\vc{A}^T \mx{G} \vc{B} = A^a G_{ab} B^b, 
\qquad\qquad
|\vc{A}| \equiv \sqrt{\vc{A}\cdot\vc{A}},
}
for any two vector fields $\vc A$ and $\vc B$. The cotangent 
vector $\vc A^\dag$ is defined by 
\(
(\vc A^\dag)_a \equiv A^b G_{ba}.
\)
The Hermitean conjugate $\mx{L}^\dag$ of a linear operator 
$\mx{L}: T_p \cM \lra T_p \cM$ 
with respect to this inner product is defined by
\equ{
\vc{B} \cdot (\mx{L}^\dag \vc{A}) \equiv
(\mx{L} \vc{B}) \cdot \vc{A},
}
so that $\mx{L}^\dag = \mx{G}^{-1} \mx{L}^T \mx{G}$.
A Hermitean operator $\mx H$ satisfies 
$\mx{H}^\dag = \mx{H}$. 
An important example of Hermitean operators are the projection 
operators. Apart from being Hermitean, a projection operator $\mx P$ 
is idempotent: $\mx{P}^2 = \mx{P}$.

To complete our discussion on the geometry of $\cM$ we
introduce different types of derivatives.
In the first place we have the covariant derivative on the manifold, 
denoted by $\nabla_a$, which acts in the usual way, i.e.\ 
\equ{
\nabla_b A^a   \equiv A^a_{\; ,b} + \gG^a_{bc} A^c
}
on a vector $A^a$. On a scalar function $V$, the 
derivative $\Bder$ and the covariant derivative $\Bnabla$ are 
equal
\(
(\Bnabla V)_a = (\Bder V)_a \equiv V_{,a}.
\)
If we represent $\d\Bgf$ as a standing vector, $\Bnabla$ and 
$\Bder$ are naturally lying vectors and therefore $\Bnabla^T$ and 
$\Bder^T$ are standing vectors. The second covariant derivative of a 
scalar function $V$ is a matrix with two lower indices:
\(
(\Bnabla^T \Bnabla V)_{ab} = \nabla_a \nabla_b V.
\)
The covariant derivative $\cD_\gm$ with respect to the spacetime variable
$x^\mu$ on a vector $\vc{A}$ of the tangent bundle is defined in components as
\equ{
\cD_\mu A^a \equiv
\der_\gm A^a + \gG^a_{bc} \der_\gm \gf^b A^c,
}
while $\cD_\gm$ acting on a scalar is simply equal to $\der_\gm$.
Notice that the spacetime derivative of the background field $\der_\mu \Bgf$
and the field perturbation $\Bgd\Bgf$ transform as vectors, even though the
fields $\Bgf$ in general do not, as they are coordinates on a manifold.

\end{document}